\input harvmac

\Title{\vbox{\baselineskip12pt\hbox{OSU-M-96-1}\hbox{DUKE-TH-96-103}%
\hbox{WIS-95-62-PH}\hbox{hep-th/9601108}}}
{\vbox{\centerline{Enhanced Gauge Symmetry in Type II String Theory}}}
\centerline{\authorfont Sheldon Katz$^a$, David R. Morrison$^b$, and
M. Ronen Plesser$^c$}
\bigskip\bigskip\centerline{%
$^a$Department of Mathematics, Oklahoma State
University, Stillwater, OK  74078, USA}
\centerline{$^b$Department of Mathematics, Duke University,
Box 90320, Durham, NC  27708-0320, USA}
\centerline{$^c$Department of Particle Physics,
Weizmann Institute of Science, Rehovot  76100, Israel}

\noblackbox

\font\sc=cmcsc10

% slanted Lambda to mirror V
\mathchardef\varLambda="0103
%\mathchardef\varDelta="0101

\def\CN{{\cal N}}
\def\cL{{\cal L}}
\def\cK{{\cal K}}
\def\cO{{\cal O}}
\def\cF{{\cal F}}
\def\IZ{{\bf Z}}
\def\IC{{\bf C}}
\def\D{\Delta}
\def\P#1#2{{{\bf P}^{#1}_{#2}}}
\def\Normal{{\cal N}}
\def\R{{\bf R}}

\lref\fhsv{S. Ferrara, J. A. Harvey, A. Strominger, and C. Vafa,
     {\it Second-Quantized Mirror Symmetry}, Phys. Lett. {\bf 361B} (1995)
     59--68.  [hep-th/9505162]}
\lref\bsvii{M. Bershadsky, V. Sadov, and C. Vafa, {\it
     $D$-Branes and Topological Field Theories}.
     [hep-th/9511222]}
\lref\stromopen{A. Strominger, {\it Open $p$-Branes}. [hep-th/9512059]}
\lref\david{D. Kutasov, {\it Orbifolds and Solitons}. [hep-th/9512145]}
\lref\boundstates{E. Witten, {\it Bound States of Strings and $p$-Branes}.
     [hep-th/9510135]}
\lref\trento{D. R. Morrison, {\it Through the Looking Glass},
     Lecture at CIRM conference, Trento (June, 1994), to appear.}
\lref\AMKthree{P. S. Aspinwall and D. R. Morrison, {\it String Theory on
     $K3$ Surfaces}, Essays on Mirror Manifolds II (B. R. Greene and
     S.-T. Yau, eds.), to appear. [hep-th/9404151]}
\lref\predictions{D. R. Morrison, {\it Making Enumerative Predictions by
     Means of Mirror Symmetry}, Essays on Mirror Manifolds II (B. R. Greene
     and S.-T. Yau, eds.), to appear. [alg-geom/9504013]}
\lref\reid{M.~Reid, {\it Canonical 3-Folds}, Journ\'ees de G\'eom\'etrie
     Alg\'ebrique d'Angers (A. Beauville, ed.), Sijthoff and Noordhoff,
     Alphen aan den Rijn, 1980, pp.~273--310.}
\lref\kodaira{K. Kodaira, {\it On Compact Analytic Surfaces, II--III},
     Annals of Math. {\bf 77} (1963) 563--626; {\bf 78} (1963) 1--40.}
\lref\HT{C. M. Hull and P. K. Townsend, {\it Unity of Superstring Dualities},
     Nucl. Phys. {\bf B438} (1995) 109--137. [hep-th/9410167]}
\lref\CDFV{A. Ceresole, R. D'Auria, S. Ferrara, and A. Van Proyen, {\it
     Duality Transformations in Supersymmetric Yang--Mills Theories
     Coupled to Supergravity}, Nucl. Phys. {\bf B444} (1995) 92--124.
     [hep-th/9502072]}
\lref\costrings{B. R. Greene, A. Shapere, C. Vafa, and S.-T. Yau,
     {\it Stringy Cosmic Strings}, Nucl. Phys. {\bf B337} (1990) 1--36.}
\lref\duff{M. Duff, {\it Strong/Weak Coupling Duality from the Dual String},
     Nucl. Phys. {\bf B442} (1995) 47--63. [hep-th/9501030]}
\lref\dynamics{E. Witten, {\it String Theory Dynamics in Various
     Dimensions},
     Nucl. Phys. {\bf B443} (1995) 85--126. [hep-th/9503124]}
\lref\HS{J. A. Harvey and A. Strominger, {\it The Heterotic String is a
     Soliton}, Nucl. Phys. {\bf B449} (1995) 535--552.
     [hep-th/9504047]}
\lref\gaugeKthree{P. S. Aspinwall, {\it  Enhanced Gauge Symmetries and $K3$
     Surfaces}, Phys. Lett. {\bf B357} (1995) 329--334. [hep-th/9507012]}
\lref\comments{E. Witten, {\it Some Comments On String Dynamics},
     Proc. Strings '95. [hep-th/9507121]}
\lref\OV{H. Ooguri and C. Vafa, {\it Two-Dimensional Black Hole
     and Singularities of CY Manifolds}. [hep-th/9511164]}
\lref\BBS{K. Becker, M. Becker, and A. Strominger, {\it
     Fivebranes, Membranes and Non-Per\-tur\-ba\-tive String Theory},
     Nucl. Phys. {\bf B456} (1995) 130--152.
     [hep-th/9507158]}
\lref\andy{A. Strominger, {\it Massless Black Holes and Conifolds in String
     Theory}, Nucl. Phys. {\bf B451} (1995) 97--109. [hep-th/9504090]}
\lref\rbat{V. V. Batyrev, {\it Dual Polyhedra and Mirror Symmetry for
     Calabi--Yau Hypersurfaces in Toric Varieties}, J. Alg. Geom. {\bf 3}
     (1994) 493--535. [alg-geom/9310003]}
\lref\GZK{I. M. Gel'fand, A. V. Zelevinski\u\i\ and M. M. Kapranov,
     {\it Hypergeometric Functions and Toral (Toric) Manifolds}, Func.
     Anal. Appl. {\bf 28}
     (1989) 94--106 (from Russian in Funk. Anal. Pril. {\bf 23}).}
\lref\ghl{C. G\'omez, R. Hern\'andez and E. L\'opez, {\it $S$-Duality
     and the Calabi--Yau Interpretation of the $N{=}4$ to $N{=}2$
     Flow}. [hep-th/9512017]}
\lref\agm{P. S. Aspinwall, B. R. Greene, and D. R. Morrison, {\it
     Calabi--Yau Moduli Space, Mirror Manifolds and Spacetime Topology
     Change in String Theory}, Nucl. Phys. {\bf B416} (1994) 414--480.
     [hep-th/9309097]}
\lref\mondiv{P. S. Aspinwall, B. R. Greene, and D. R. Morrison, {\it
     The Monomial-Divisor Mirror Map}, Internat. Math. Res. Notices
     (1993) 319--337. [alg-geom/9309007]}
\lref\mirrorII{D. R. Morrison, {\it Mirror Symmetry and the Type II
     String}, Proc. Trieste Workshop on $S$-Duality and Mirror Symmetry.
     [hep-th/9512016]}
\lref\udual{P. S. Aspinwall and D. R. Morrison, {\it $U$-Duality and
     Integral Structures}, Phys. Lett. {\bf 355B} (1995) 141--149.
     [hep-th/9505025]}
\lref\MorrisonPlesser{D. R. Morrison and M. R. Plesser, {\it Summing
     the Instantons: Quantum Cohomology and Mirror Symmetry in Toric
     Varieties}, Nucl. Phys. {\bf B440} (1995) 279--354. [hep-th/9412236]}
\lref\zerger{T. E. Zerger, {\it Contracting Rational Curves on Smooth
     Complex Threefolds}, Ph.D. thesis, Oklahoma State University, 1996.}
\lref\gms{B. R. Greene, D. R. Morrison, and A. Strominger,
     {\it Black Hole Condensation and the Unification of String Vacua},
     Nucl. Phys. {\bf B451} (1995) 109--120. [hep-th/9504145]}
\lref\twoparamsI{P. Candelas, X. de la Ossa, A. Font, S. Katz, and
     D. R. Morrison, {\it Mirror Symmetry for Two Parameter Models (I)},
     Nucl. Phys. {\bf B416} (1994) 481--562. [hep-th/9308083]}
\lref\phases{E. Witten, {\it Phases of $N{=}2$ Theories In Two Dimensions},
     Nucl. Phys. {\bf B403} (1993) 159--222. [hep-th/9301042]}
\lref\rCG{C. H.~Clemens and P. A.~Griffiths, {\it The Intermediate Jacobian
     of the Cubic Threefold}, Annals of Math. {\bf 95} (1972) 281--356.}
\lref\rBour{N.~Bourbaki, {\it Groupes et alg{\`e}bres de Lie\/}, Hermann,
     Paris, 1968.}
\lref\BCOV{M. Bershadsky, S. Cecotti, H. Ooguri, and C. Vafa, {\it Holomorphic
     Anomalies in Topological Field Theories} (with an appendix by S. Katz),
     Nucl. Phys. {\bf B405} (1993) 279--304. [hep-th/9302103]}
\lref\rbkkI{P.~Berglund, S.~Katz, and A.~Klemm, {\it Mirror Symmetry and the
     Moduli Space for Generic Hypersurfaces in
     Toric Varieties}, Nucl. Phys. {\bf B456} (1995) 153--204.
     [hep-th/9506091]}
\lref\HKTYI{S. Hosono, A. Klemm, S. Theisen, and
     S.-T. Yau, {\it Mirror Symmetry, Mirror Map and Applications to
     Calabi--Yau
     Hypersurfaces}, Commun. Math. Phys. {\bf 167} (1995) 301--350.
     [hep-th/9308122]}
\lref\HKTYII{S. Hosono, A. Klemm, S. Theisen, and
     S.-T. Yau, {\it Mirror Symmetry, Mirror Map and Applications to Complete
     Intersection Calabi--Yau Spaces}, Nucl. Phys. {\bf B433} (1995) 501--554.
     [hep-th/9406055]}
\lref\rKKLMV{S. Kachru,  A. Klemm, W. Lerche, P. Mayr and C. Vafa,
     {\it Non-Perturbative Results on the Point Particle Limit
     of $N{=}2$ Heterotic String Compactifications}, Nucl. Phys. {\bf B459}
     (1996) 537--588. [hep-th/9508155]}
\lref\paul{P. S. Aspinwall, {\it Enhanced Gauge Symmetries and Calabi--Yau
     Threefolds}. [hep-th/9511171]}
\lref\rbkkII{P.~Berglund, S.~Katz, and A.~Klemm, {\it Extremal Transitions
     between Dual $N{=}2$ String Models\/}, in preparation.}
\lref\bsvi{M. Bershadsky, V. Sadov, and C. Vafa, {\it $D$-Strings on
     $D$-Manifolds}. [hep-th/9510225]}
\lref\rKap{M. M. Kapranov, {\it A Characterization of $A$-Discriminantal
     Hypersurfaces
     in Terms of the Logarithmic Gauss Map}, Math. Ann. {\bf 290} (1991)
     277--285.}
\lref\kv{S. Kachru and C. Vafa, {\it Exact Results for $N{=}2$
     Compactifications of Heterotic Strings}, Nucl. Phys. {\bf B450} (1995)
     69--89. [hep-th/9505105]}
\lref\vafa{C.~Vafa, {\it A Stringy Test of the Fate of the Conifold},
     Nucl. Phys. {\bf B447} (1995) 252--260. [hep-th/9505023]}
\lref\NS{Y. Namikawa and J. H. M. Steenbrink, {\it Global Smoothing
     of Calabi--Yau Threefolds}, Inv. Math. {\bf 122} (1995) 403--419.}
\lref\grossdef{M.~Gross, {\it Deforming Calabi--Yau Threefolds}.
     [alg-geom/9506022]}
\lref\rgross{M.~Gross, {\it Primitive Calabi--Yau Threefolds}.
     [alg-geom/9512002]}
\lref\rnati{N. Seiberg, {\it Supersymmetry and Non-Perturbative Beta
     Functions}, Phys. Lett. {\bf 206B} (1988) 75--80.}
\lref\rStrom{A. Strominger, {\it Special Geometry}, Commun. Math. Phys.
     {\bf 133} (1990) 163--180.}
\lref\sw{N. Seiberg and E. Witten, {\it Electromagnetic Duality,
     Monopole Condensation and Confinement in $N{=}2$ Supersymmetric
     Yang--Mills Theory},
     Nucl. Phys. {\bf B426} (1994) 19--52. [hep-th/9407087]}
\lref\beyond{D. R. Morrison, {\it Beyond the {K\"a}hler Cone},
     Proc. of the Hirzebruch 65 Conference on
     Algebraic Geometry (M.~Teicher, ed.), Israel Math. Conf. Proc., vol.~9,
     Bar-Ilan University, 1996, pp.~361--376. [alg-geom/9407007]}
\lref\small{
     P. S. Aspinwall, B. R. Greene and D. R. Morrison, {\it Measuring Small
     Distances in
     $N{=}2$ Sigma Models}, Nucl. Phys.  {\bf B420} (1994) 184--242.
     [hep-th/9311042]}
\lref\pauljan{P. S. Aspinwall and J. Louis, {\it On the Ubiquity of $K3$
     Fibrations in String Duality}. [hep-th/9510234]}
\lref\tft{E. Witten, {\it Topological Quantum Field Theory},
     Commun. Math. Phys. {\bf 117} (1988) 353--386.}
\lref\chs{C. Callan, J. Harvey and A. Strominger, {\it Supersymmetric
     String Solitons}, Proc. of 1991 Trieste Spring School, String
     Theory and Quantum Gravity (1991) 208--244. [hep-th/9112030]}
\lref\cdgp{P.~Candelas, X.~C. de~la Ossa, P.~S. Green, and L.~Parkes,
     {\it A Pair of
     Calabi--Yau Manifolds as an Exactly Soluble Superconformal Theory},
     Nucl. Phys.  {\bf B359} (1991) 21--74.}
\lref\tftrc{P. S. Aspinwall and D. R. Morrison, {\it Topological Field
     Theory and Rational Curves}, Commun. Math. Phys. {\bf 151} (1993)
     245--262.}
\lref\hayakawa{Y. Hayakawa, {\it Degeneration of Calabi--Yau Manifold with
     Weil--Petersson Metric}. [alg-geom/9507016]}
\lref\BR{ D.~Burns, Jr. and M.~Rapoport, {\it
     On the {T}orelli Problem for {K\"a}hlerian $K3$ Surfaces},
     Ann. Sci. {\'E}cole Norm. Sup. (4) {\bf 8} (1975) 235--274.}
\lref\Wilson{P. M. H.~Wilson, {\it The K\"ahler Cone on Calabi--Yau
     Threefolds},
     Inv. Math. {\bf 107} (1992) 561--583.}
\lref\AspinExtreme{P. S. Aspinwall, {\it An $N{=}2$ Dual
     Pair and a Phase Transition}, Nucl. Phys. {\bf B460} (1996) 57--76.
     [hep-th/9510142]}
\lref\GimPol{E. G. Gimon and J. Polchinski, {\it
     Consistency Conditions for Orientifolds and $D$-Manifolds}.
     [hep-th/9601038]}
\lref\km{A. Klemm and P. Mayr, {\it Strong Coupling Singularities
     and Non-Abelian Gauge Symmetries in $N{=}2$ String Theory}.
     [hep-th/9601014]}

\vskip.3in

We show how enhanced gauge symmetry in type II string theory compactified on a
Calabi--Yau threefold arises from singularities in the geometry of the target
space.  When the target space of the type IIA string acquires a genus $g$ curve
$C$ of $A_{N-1}$ singularities, we find that an $SU(N)$ gauge theory with $g$
adjoint hypermultiplets appears at the singularity.  The new massless states
correspond to solitons wrapped about the collapsing cycles, and their dynamics
is described by a twisted supersymmetric gauge theory on $C\times \R^4$.  We
reproduce this result from an analysis of the $S$-dual $D$-manifold.  We check
that the predictions made by this model about the nature of the Higgs branch,
the monodromy of period integrals, and the asymptotics of the one-loop
topological amplitude are in agreement with geometrical computations.  In one
of our examples we find that the singularity occurs at strong coupling in the
heterotic dual proposed by Kachru and Vafa.

\Date{1/96; revised 3/96}

\newsec{Introduction}

Recent progress in string theory has involved the application of
new techniques to study nonperturbative effects.
These include string-string duality which relates heterotic
to type II string theory. This duality is quite well understood
in six-dimensional
examples (and in
$\CN{=}4$ supersymmetric models in four dimensions obtained from these)
where it relates strong to weak coupling
\refs{\HT\CDFV\dynamics\HS\gaugeKthree\comments{--}\david}.
There are in addition a number of examples of $\CN{=}2$ models in four
dimensions for which conjectured dual formulations have been proposed
\refs{\kv,\fhsv}.
Much less is understood about this duality in four dimensions, but
it clearly requires that nonabelian gauge symmetry appear at special
points in the type II moduli space \refs{\rKKLMV\AspinExtreme\bsvi{--}\paul}.

Nonperturbative physics has been probed in another context in the
study of singularities in the moduli
space of perturbative vacua of type II string theory.  In particular,
for vacua realized as Calabi--Yau compactifications, the singularities
in moduli space where the target space acquires conifold singularities
have been understood as resulting from the existence of solitonic
states whose mass vanishes at the singular points \andy.
These states are in hypermultiplets of the $\CN{=}2$ algebra in four
dimensions, and are charged under the abelian  symmetry gauged by the
Ramond-Ramond vector fields.  When the spectrum of massless states
is such that the moduli space of supersymmetric vacua in the gauge
theory contains a Higgs branch, the transition to the Higgs phase
has been interpreted as describing a
topology-changing transition to the moduli space of another
Calabi--Yau manifold \gms.
In this work, we study the singularities in Calabi--Yau moduli space
which give rise to nonabelian gauge symmetry.  As anticipated in
\refs{\AspinExtreme\bsvi{--}\paul},
such enhanced gauge symmetry is associated to
the presence of {\it curves of singularities}\/ $C$ on the Calabi--Yau
manifold.  We study the spectrum and dynamics of the massless solitons
in this case, exhibiting the origins of the enhanced symmetry.

In order to stay at finite distance in the Calabi--Yau moduli space,
we must restrict our attention to singular Calabi--Yau spaces which
can be resolved by a Calabi--Yau manifold \hayakawa, in other words,
the Calabi--Yau space should have only so-called {\it canonical
singularities}. It has long been known in the mathematics literature \reid\
that at the generic point of a curve of canonical
singularities on a Calabi--Yau
space, the transverse singularity
is a quotient singularity $\IC^2/G$ (which can be resolved by an ALE
space).  We consider in this paper
the case in which the curve of singularities is smooth and irreducible,
the singularity type is constant
along the entire curve, and $G=\IZ/N\IZ$ so that the singularity
is of type $A_{N-1}$. An example with two intersecting curves of singularities
was discussed (from a different point of view) in \bsvi; examples in which
the singularity type is allowed
to change along the curve are currently under study \rbkkII.

Concretely, we study type IIA
models with a singularity in codimension $N{-}1$ in
moduli space at which the manifold acquires a genus $g$ curve of
$A_{N-1}$ singularities.  The singularity can be smoothed to obtain a
smooth, topologically distinct manifold generalizing the conifold
transitions studied in \gms.  We find that in this case the
four-dimensional effective field theory describing the low-energy dynamics
is an $\CN{=}2$ gauge theory with gauge group $SU(N)$ and $g$
hypermultiplets transforming in the adjoint representation.  We make
some assumptions in obtaining this result, and these are verified by
a detailed comparison of the predictions of this model to geometrical
computations in a list of examples.  These include the local geometry
of the moduli space, the properties of the Higgs branch (moduli space
of the manifold obtained after smoothing), monodromy of period integrals
(in the type IIB representation of the model using mirror symmetry)
and asymptotic behavior of the one-loop topological amplitude.

We restrict our attention to the case $g{>}1$.
In \paul\ the case of a degeneration in the fibers of a $K3$ fibration
of a Calabi--Yau space (in which each fiber acquires a single $A_{N-1}$
singularity) was considered.  This is a special case of a
{\it rational}\/ curve of singularities (i.e., $g{=}0$)\foot{As Aspinwall
has pointed out to us, it may be possible to find examples with $g{>}0$
using his construction by relaxing the requirement imposed in \paul\
of trivial monodromy of the Picard lattice.};
the enhanced gauge symmetry was observed in the limit of
infinite area for the singular curve. This is consistent with the
extension of our result to this case.  The pure $\CN{=}2$ Yang--Mills
theory is asymptotically free and the nonabelian symmetry is only
restored in the extreme high-energy (weak-coupling) limit. In an
example this was directly observed in the dual heterotic theory in
\refs{\kv,\rKKLMV}.
The other special case $g{=}1$ leads, according to our rule of $g$ matter
hypermultiplets, to the $\CN{=}4$ theory in four dimensions.  In this
(conformal) model the nonabelian gauge symmetry is restored at the
origin of the Coulomb branch, where the low-energy physics is
described by a nontrivial interacting conformal field theory; the
Higgs branch is absent.  In \bsvi\ the case of a degeneration in the
$K3$ factor of a $K3\times T^2$ compactification (which has $\CN{=}4$
supersymmetry at generic points) was shown to conform to this
prediction.  Our analysis suggests that the singular behavior depends
only on the local form of the singularity and hence the results for
these special cases should extend more generally.

One of the models we study in detail has a conjectured heterotic dual \kv.
It is
interesting that the singularity we study occurs in the strong
coupling region of the heterotic model.\foot{The locus in moduli space
at which we are finding enhanced
gauge symmetry at strong (heterotic) coupling
is completely disjoint from the locus of
weak-coupling enhanced gauge symmetry discussed in
\refs{\kv,\rKKLMV,\paul}.}
Thus the enhanced symmetry is
presumably a nonperturbative effect in the heterotic model as well,
distinct from the well-known appearance of enhanced symmetries in the
conformal field theory.  It would be extremely interesting to
understand the phenomenon from the point of view of the heterotic
theory.

The paper is organized as follows. In section two we study the
dynamics of the light solitons near the singular point in moduli space.
The essential pont
is that the topology of the two-cycles whose area shrinks to zero at
the singularity is completely
determined by the singularity in the space transverse to $C$.  Locally
on $C$, we thus have a situation similar to the case of an $A_{N-1}$
point on $K3$, leading to an enhanced gauge symmetry in an effective
six-dimensional theory.  In our case, the light states
propagate in a six-dimensional spacetime of the form $C\times
\R^4$, and their dynamics is governed by a twisted
supersymmetric gauge theory on this space.  The twisting is
essentially determined by the requirement of
$\CN{=}2$ supersymmetry in the effective four-dimensional theory.
We use this model to
make some detailed predictions about the geometry of the Calabi--Yau
moduli space near the singularity.  These are checked by geometrical
computations in the next section, and we find complete agreement.
We find a dual
description of this in terms of $D$-strings on a $D$-manifold, refining
the proposal of \refs{\bsvi,\OV}. In this context the twisting we
found is related to that studied in \bsvii .

In section three, we study the local geometry of the Calabi--Yau space
and the moduli space near the singularity, verifying the predictions
of the previous section in detail.
We observe the expected action of the Weyl group of $A_{N-1}$ for both the
type IIA and type IIB models and also see the structure of the $A_{N-1}$ root
lattice geometrically in several different forms.  For the type IIB model,
$A_{N-1}$ enters into the description in essentially the same way that it
enters into the deformation theory of $A_{N-1}$ singularities.  The
moduli space of conformal field theories fails to be a product of
the complex and K\"ahler moduli spaces.  (One must instead take the
quotient of a product by a finite group.)  For the
type IIA model, the key ingredient in the analysis is the notion of an
elementary transformation, which consists of an involution on the second
cohomology of the target manifold with properties similar to those of a
flop.  We match the geometry of certain
extremal transitions with the properties of the branches of our gauge
theoretic model.  We also check the monodromy and one-loop amplitude
against our predictions in several examples, and find complete agreement.

In section four we describe our detailed calculations in
two specific examples.  These are the two-parameter
model of \kv , which is of
interest because of the existence of a conjectured heterotic dual, and
a model exhibiting $SU(3)$ gauge symmetry at the transition.

\newsec{Soliton Dynamics}

The type II string theories which we consider have two geometric
representations: as a IIA theory on a certain Calabi--Yau manifold
$M$, or as a IIB theory on its mirror partner $W$.  (These two
representations are believed to be equivalent even when nonperturbative
effects are taken into account \refs{\andy,\udual,\mirrorII}.)
We will use both models interchangeably.  The low-energy theory in
four dimensions is $\CN{=}2$ supersymmetric; the
spectrum at generic points in the moduli space contains
$n_V{=}h^{21}(W){=}h^{11}(M)$ massless vector multiplets and
$n_H{=}h^{11}(W)+1{=}h^{21}(M)+1$ massless hypermultiplets.  The
moduli space is parameterized by the expectation values of the scalars
in these multiplets.  Since the dilaton is in a hypermultiplet, the
nonrenormalization theorem of \refs{\andy,\BBS} guarantees that the
geometry of
the moduli space of the scalars in the vector multiplets can be
computed exactly at string tree level.  These moduli correspond
geometrically
to complex structure deformations of $W$, or equivalently to
(complexified) K\"ahler deformations of $M$.

{}From the IIA perspective, a singularity in the conformal field theory
will occur if the K\"ahler class approaches a face of the K\"ahler
cone, with the $B$-field taking an appropriate value.\foot{The fact
that this singularity can be avoided by judicious choice of the
$B$-field was employed in \refs{\phases,\agm} to argue that conformal
field theory could be extended beyond the boundary set by the K\"ahler
cone.  However, here we must consider what happens when the singularity
is approached, not avoided.}  Each face of the K\"ahler cone (we have
in mind faces which may be of high codimension) determines a collection
of holomorphic $2$-spheres whose area shrinks to zero as the face
is approached.  These $2$-spheres can be contracted to points at the
expense of introducing singularities into the new
space.\foot{Actually, because of worldsheet instanton effects the
singularity  will occur at a finite area for the $2$-spheres.}  It was
Strominger's insight \andy\ that nonperturbative effects will ``cancel''
these singularities, leading to nonsingular physics.

A simple case to consider is the one in which only a finite number of
$2$-spheres on $M$
are contracted by this process---this is the case studied
in \refs{\andy,\gms} (albeit from the IIB perspective).
Another simple case is one in which a divisor
is contracted to a smooth curve.  In this paper we consider a slight
generalization of that second case, in which a collection of $N{-}1$
divisors is
contracted to a smooth curve of genus $g$.  We assume that the
singularities
along the curve are uniformly of type $A_{N-1}$.\foot{There is evidence that
similar results apply if the singularities are more complicated
at finitely many points of the curve \rbkkII.}  In this
section we seek to understand the properties of the light solitonic
states in the vicinity of this singularity.

The transverse $A_{N-1}$ singularity is resolved by an ALE space.
The  cycles which shrink to zero area
are described by a chain of $N-1$ two-spheres
$\Gamma_i$ in this space, with their intersection matrix corresponding
to the Dynkin diagram of $A_{N-1}$.  As we move about $C$ these
spheres sweep out $N{-}1$ divisors $E_i$ on $M$.
In homology there are then $(N{-}1)$ shrinking
two-cycles $\Gamma_i$, and $N{-}1$ shrinking four-cycles $E_i$.
The soliton states we
consider are described by two-branes wrapping about the two-cycles
$A^{ij}$ for $i{\leq }j$ represented
by the chain $\Gamma_i\cup\Gamma_{i+1}\cup\cdots\cup\Gamma_j$.
(We call these the {\it collapsing cycles}.)
Under the $U(1)^{N-1}$
symmetry gauged by the Ramond-Ramond superpartners of the moduli
corresponding to $E_i$, their charges correspond to the positive roots
of $A_{N-1}$; the negative roots correspond to
orientation-reversed wrappings.
\foot{There are additional massless
solitons from four-branes wrapping around the shrinking four-cycles
$A_D$. These are magnetically charged states.  As we shall see,
world-sheet instantons correct their masses, leading to agreement with
the monopole masses in the gauge theory description.  These should not
be considered independent degrees of freedom in the low-energy
description. We thank N.~Seiberg for a helpful comment.}

\subsec{The Soliton Spectrum}

The massless states associated to two-branes wrapping
around $A$ are in some sense intermediate between two cases that have
recently been studied.  In \refs{\andy,\gms} the case of isolated
rational curves in a Calabi--Yau space was studied, and these were
found to lead to massless hypermultiplets in the effective
four-dimensional $\CN{=}2$ field theory, with $U(1)$ charges
determined by their homology class.  Our case differs from this in
that we have a two-dimensional moduli space of such curves (given by
$C$).  In \refs{\dynamics,\gaugeKthree}
the case of rational curves on a
$K3$ manifold was studied. Here the particle states were found to fill
out a massive vector multiplet of the $\CN{=}(1,1)$ supersymmetric theory
in six dimensions with gauge group $SU(N)$.  The gauge indices arise,
as described in the previous paragraph, from the labeling of the
two-cycles.  Our models are similar to this in that there is a
separation of scales between the size of $\Gamma_i$ and the size of
$C$, which can
be arbitrarily large since the size of $C$ is a modulus of the
singular locus.  Thus we can approximately think of the low-energy
theory as a compactification on $C$ of the six-dimensional
theory obtained by including the massless solitons wrapping around
$A^{ij}$.

If we adopt this picture too na\"{\i}vely we will of course run into
difficulties.  The compactification on $C\times \R^4$ will not be
supersymmetric because $C$ has no covariantly
constant spinor.  The supersymmetry that acts in six dimensions must
be {\it twisted}, as
recently proposed in \bsvii\ in a different context.  Heuristically,
we can think of the twisting as follows.  The approximation in which
we find the six-dimensional theory relies on the existence of a
neighborhood of $E_i$ in $M$ which fibers over $C$.  The fibration
however is nontrivial, and this must in some way predict the
twisting. We do not know how to compute this effect directly, but as we shall
see the twist is almost uniquely determined by the requirement that
the four-dimensional effective theory exhibit $\CN{=}2$
supersymmetry.

In flat space the six-dimensional theory contains a vector
$V_M$, a complex scalar $\phi$, and two fermions, all
in the adjoint representation of $SU(N)$.  The charged components are
the soliton states; the $N{-}1$ neutral components are supplied by the
$N{-}1$ moduli of the ALE space.  The fields transform under a global
$SU(2)\times SU(2)$ $R$-symmetry.
The compactification breaks the local Lorentz group as $SO(6)\to
SO(4)\times U(1)$, and the reduction leads to the following
representations carried by the fields and the supercharges
\eqn\reps{
\matrix{
&SO(4)&\times &U(1)&\times &SU(2)&\times &SU(2)\cr
V_\mu&{\bf (2,2)}&&0&&{\bf 1}&&{\bf 1}\cr
V_{++}&{\bf (1,1)}&&1&&{\bf 1}&&{\bf 1}\cr
V_{--}&{\bf (1,1)}&&-1&&{\bf 1}&&{\bf 1}\cr
\phi&{\bf (1,1)}&&0&&{\bf 2}&&{\bf 2}\cr
\psi&{\bf (1,2)}&&1/2&&{\bf 2}&&{\bf 1}\cr
\lambda&{\bf (1,2)}&&-1/2&&{\bf 1}&&{\bf 2}\cr
Q_\alpha&{\bf (1,2)}&&1/2&&{\bf 1}&&{\bf 2}\cr
\bar Q_\alpha&{\bf (1,2)}&&-1/2&&{\bf 2}&&{\bf 1}\cr
}}
where boldface numbers give the dimension of the representation, and
$SO(4)$ representations are labeled using the local isomorphism
$SO(4)\sim SU(2)\times SU(2)$.

A twist \tft\ is a choice of an exotic action of the
$U(1)$ local Lorentz transformation group of $C$, as a subgroup
$U(1)'$ of $U(1)\times SU(2)\times SU(2)$.  The unbroken supersymmetry
is generated by the supercharges that are scalars under this action.
We expect the twisted theory to exhibit $\CN{=}2$ supersymmetry in four
dimensions, and this requirement leaves two possible twists.  The one
we will use (and we justify eliminating the other by the results) is
\eqn\twist{
J' = J_L-J_3^{(1)}-J_3^{(2)}\ ,
}
where $J_L$ is the ``standard'' Lorentz generator, and the other two
correspond to  the Cartan elements of the $SU(2)$'s. In four
dimensions the global symmetry is the subgroup of $U(1)\times
SU(2)\times SU(2)$ which commutes with the holonomy (in other words,
with $U(1)'$).  This will give the $SU(2)_R\times U(1)$ $R$-symmetry
of the four-dimensional theory.

The massless
modes in four dimensions arise from zero modes of the relevant
differential operators on $C$.  Thus states with $U(1)'$ charge zero
lead to one
massless particle in four dimensions, corresponding to the unique
(constant) holomorphic function on $C$.
These comprise an $\CN{=}2$
vector multiplet, including $V_\mu$, a complex scalar made from the two
neutral components of $\phi$, and the neutral components of the
fermions.  States with charge $\pm 1$ lead to a massless particle for
each (anti-) holomorphic 1-form on $C$, thus to precisely $g$
particles.  These comprise $g$ hypermultiplets including all of the
remaining fields.

We can see here the two special cases $g{\leq}1$.  For $g{=}0$, there
are no massless hypermultiplets and we find the pure Yang--Mills theory
in four dimensions \paul .  For $g{=}1$ there is no holonomy and we
will find the full $SU(4)$ $R$-symmetry of four-dimensional $\CN{=}4$
gauge theory in four dimensions, though only a maximal subgroup is
manifest in \reps . In this case the magnetically charged states
corresponding to four-branes wrapping four-cycles do not receive
logarithmic corrections to their masses.  The
$\CN{=}4$ theory has an electric-magnetic duality exchanging
electrically charged states with magnetically charged ones.  In our
case this is manifestly accomplished by $\rho\to-1/\rho$ where $\rho$
is the K\" ahler modulus of the torus.  We thus see the expected
relation of $T$-duality for the IIB theory with $S$-duality of the
heterotic dual \refs{\duff,\dynamics}.

In general we predict that the low-energy theory near the singularity will
be described by four-dimensional $\CN{=}2$ supersymmetric gauge theory
with gauge group $SU(N)$ and $g$ hypermultiplets of matter in the
adjoint representation.  The other possible twist mentioned above is easily
seen to lead in four dimensions to pure Yang--Mills theory.  As we
shall see in the next section, this does not describe the singularity
correctly.  It would be very interesting to compute the twisting
directly, as was done in \bsvii\ for the situation studied there.
In both of the special cases we saw explicitly that the modulus of the
IIA theory corresponding to the area of $C$ represents the gauge
coupling in the effective field theory.  It is natural to assume that
this holds for $g{>}1$ as well.

\subsec{Predicting the Local Geometry}

We have asserted that at
the singularity in question the spectrum of massless states is the
same as that of $\CN{=}2$ supersymmetric $SU(N)$ gauge theory with $g$
hypermultiplets transforming in the adjoint representation of the
gauge group.  The structure of the moduli space predicted by this
model in the vicinity of the singularity can be reliably computed at
the classical level, since the theory is not asymptotically free (for
$g>1$).

In
$\CN{=}1$ superfield notation, the massless fields in the theory are
the (real) vector multiplet in the adjoint $V_a$, with
field strength $W_\alpha^a$; a (complex) chiral superfield in the
adjoint $\Phi^a$ (comprising the vector multiplet described above);
and $2g$ chiral superfields $M^i_a$ and ${\tilde M}^i_a$ in the adjoint
(comprising the $g$ hypermultiplets), where $i$ runs from $1$ to $g$.

Under the $SU(2)_R$ symmetry discussed in the previous subsection the
scalars $m$ and $\tilde m$ are a doublet, while
the bosons in $V$ and $\Phi$ are invariant.  The Lagrangian is
\eqn\el{\eqalign{
2\pi {\cal L} &= {\rm Im}\left[ {\rm Tr} \int d^4\theta
\left( M^{\dagger}_i e^V M^i + {\tilde M}^{\dagger i}e^V{\tilde M}_i
+\Phi ^\dagger e^V\Phi\right)\right.\cr
&\qquad\left.+{\tau\over 2}\int d^2\theta{\rm Tr} W^2 +
i\int d^2\theta {\cal W}\right]\cr}
}
where the superpotential is
\eqn\ew{
{\cal W} = {\rm Tr} {\tilde M}^i [\Phi,M_i]\ .
}
The scalar potential is
\eqn\ev{\eqalign{
{\cal V} &= {\rm Tr}\left[ [m_i,m^{\dagger i}]^2
+[{\tilde m}^i,{\tilde m}^\dagger_i]^2+[\phi,\phi^\dagger ]^2\right.\cr
&\left.+2\left( [m^{\dagger i},\phi][\phi^\dagger,m_i]+
[{\tilde m}^\dagger_i,\phi][\phi^\dagger,{\tilde m}^i]
+[m_i,{\tilde m}^i][{\tilde m}^\dagger_j,m^{\dagger j}]
\right)\right]\ .\cr}
}
Each of the summands is of the form ${\rm Tr} AA^\dagger$,
thus each must vanish separately as the condition for a supersymmetric
vacuum.

The moduli space of supersymmetric vacua contains a Coulomb branch,
along which $\phi$ acquires an expectation value. From \ev\ this must
satisfy $[\phi,\phi^\dagger]=0$, which means it can be brought by a gauge
transformation to the form $\phi = {\rm
diag}(\phi_1,\phi_2,\ldots,\phi_N)$ subject to the constraint $\sum
\phi_i=0$.  The $\phi_i,\,i{=}1,\ldots,N{-}1$ serve as coordinates on
an $S_N$ cover of the Coulomb branch, since the Weyl group acts by permuting
the eigenvalues (leading to an action on the coordinates by the
$(N{-})1$-dimensional representation).  At generic points along this the gauge
symmetry is spontaneously broken to $U(1)^{N-1}$.  The massless
components of the matter hypermultiplets are the diagonal elements in
this basis, as is evident from \ev . There are thus $g(N{-}1)$ neutral
massless hypermultiplets.  The Weyl group acts on these via the
$(N{-}1)$-dimensional representation mentioned above.
Since the moduli space is a quotient by the action of this group
the factorization into hypermultiplet moduli $\times$ vector moduli
will fail at the origin. This local quotient structure of the moduli
space can be taken as the zeroth prediction from our model.
In the remainder of this subsection we
will use the identification of the singularity as a point of enhanced
gauge symmetry to make additional predictions which will be
verified in section three by a closer look at the geometry.

The first prediction is the existence of a Higgs branch.  As in \gms ,
the scalar potential allows the solitons to condense, and following
the example there we anticipate that this will describe a
generalization of the conifold transition to a distinct Calabi--Yau
manifold.  Along this branch the scalars in the hypermultiplets
acquire expectation values, and at a generic point the gauge group is
completely broken.  The (quaternionic) dimension of the Higgs branch
is predicted to be
\eqn\ehigs{
{\cal H} = (g{-}1)(N^2{-}1)\ .
}

We can somewhat refine this prediction.
There are special submanifolds on the Coulomb branch (meeting at the
origin) along which nonabelian subgroups of $SU(N)$ are unbroken.  These
correspond exactly to fixed point sets of the $S_N$ action,
where eigenvalues of $\phi$ coincide. In general an unbroken symmetry
\eqn\unbroken{
SU(k_1)\times\cdots\times SU(k_p)\times U(1)^{p-1}
}
where $\sum k_i = N,\,k_i\geq 1$ (and factors
of $SU(1)$ are simply to be ignored) will occur in codimension
$N{-}p$, and there will be $g$ massless hypermultiplets in the adjoint
representation of the unbroken group.  We can allow these to acquire
expectation values, breaking the nonabelian part completely and
leading to a mixed branch with the Higgs component having dimension
\eqn\emix{
(g-1)\sum_{i=1}^p (k_i^2-1)+g(p-1)\ .
}

We can also see the transition to the Higgs branch along a different
path which will prove more transparent in the geometrical analysis.
At a generic point on the Coulomb branch, we can turn on expectation
values for the $g(N{-}1)$ neutral hypermultiplet scalars.  Then, as we
tune $\phi$ to zero, the nonabelian symmetry is not restored, and the
gauge symmetry remains $U(1)^{N-1}$, as in \gms .  The theory is still
IR free and we can use classical analysis.  From \ev\ we see
that the hypermultiplet expectation values lead to masses for the
off-diagonal components with rank $N^2{-}N$ (essentially  there is a
mass for the off-diagonal entries collinear in flavor
space with the expectation value).  Thus there are at the singular
point an additional $(g{-}1)(N^2{-}N)$ charged hypermultiplets with
the $U(1)^{N-1}$ charges of $g{-}1$ adjoints.  When these acquire
generic expectation values the gauge symmetry is Higgsed leading back
to the Higgs branch dimension \ehigs .

In codimension one (along components of the singular divisor) the
unbroken symmetry is $SU(2)\times U(1)^{N-2}$. This locus is fixed by
a $\IZ_2$ subgroup of $S_N$ and in terms of an invariant transverse
coordinate $u$ we can predict from the one-loop beta function a
logarithmic term in the effective action.  Since the metric on moduli
space is given by two derivatives of a holomorphic prepotential both
in the gauge theory \rnati\ and in the conformal field theory \rStrom\
(but note that here we are going to neglect gravitational effects and
use the form of the K\"ahler potential appropriate to global
supersymmetry, see \sw ) we identify the singular terms of these.
In both theories the prepotential is given by identical expressions in
terms of a symplectic basis of periods.  In the gauge theory case
these are the periods of an elliptic curve constructed as in \sw , while
in the conformal field theory they are the periods of the holomorphic
three-form on $W$.  Our second prediction is thus that two of the
periods of $W$
will be given in the vicinity of the singular point by the (one-loop)
expressions for the periods of the elliptic curve
\eqn\eas{\eqalign{
a &= \sqrt{2u}+\ldots\cr
a_D &= i{2-2g\over\pi}a\log a .\cr}
}
In other words, we predict that
there should be a two-dimensional subspace of shrinking cycles,
and under $u\to e^{2\pi i}u$ the monodromy in this subspace should be
given by
\eqn\emon{\pmatrix{a_D\cr a\cr}\to
\pmatrix{-1&2(g-1)\cr0&-1\cr}\pmatrix{a_D\cr a\cr}
\ .}
The second of \eas\ exhibits the logarithmic corrections to the
monopole masses, vanishing for $g{=}1$, which were discussed above.

Using the results of \vafa\ we can extend our predictions to
subleading order in the string coupling of the type II theory.
This is because there is a topological (hence computable) one-loop
amplitude which  appears as the coefficient of a particular term in
the spacetime effective action.
As Vafa showed, the singular behavior of this coefficient near a
singularity associated with the existence of new massless particle states
is given by
\eqn\evafa{
F_1^{\rm top}\sim -{1\over 12}\sum (-)^{s(i)} \log \mu_i^2
}
where the sum extends over BPS states with mass $\mu_i$
becoming massless at the
singularity and $s(i)$ is zero for hypermultiplets and one for vector
multiplets.  In the codimension-one case considered above,
the masses of the charged states are easily seen from \ev\ to be
$\mu = a$. (This follows in general from the central charge
formula but in this IR free case the classical computation is sufficient.)
There are $2g$ hypermultiplets of these, and 2 vector multiplets, so
substituting \eas\ in \evafa\ we have our third prediction
\eqn\efi{
F_1^{\rm top}\sim -{1\over 12} (g-1) \log |u|^2\ .
}

There are thus a number of detailed predictions that our model for the
nature of the singularity makes regarding the geometrical data of
$W$.  In sections three and four we will see that these are in excellent
agreement with geometrical calculations.

\subsec{$D$-String Interpretation}

Following recent discussions of string duality we can try to
understand this model from a
different point of view.  It has been argued
\refs{\comments,\bsvi} that type IIA strings on a $K3$ manifold in
the vicinity of an
$A_{N-1}$ singularity can be locally described by type IIB strings
with $N$ coalescing NS-NS fivebranes.  These are localized at the
singular points so their worldvolumes fill spacetime.  The
coalescing fivebranes lead to enhanced gauge symmetry on their
worldvolume. (This was found for $D$-branes in \boundstates , and \bsvi\
noted that this is $S$-dual to the situation at hand.  In the NS-NS
case the string configurations responsible for the enhanced gauge
symmetry were described in \stromopen .)  The worldvolume fills
spacetime, so we observe enhanced gauge symmetry in the
six-dimensional effective theory.

This model of degenerating surfaces is a bit subtle, so it is
worthwhile pausing here to review and clarify the argument.  This is
most directly done along the lines of \costrings . A $K3$ surface
can be constructed as an elliptic fibration, i.e.\ as a compactification of the
total space of a $T^2$ bundle over $S^2$,
provided that the moduli of the $K3$
are suitably restricted.\foot{If we assume
that the elliptic fibration admits a section, then these restricted
moduli comprise
18 complex-structure moduli and 2 K\"ahler moduli, the latter being
the sizes of the base and the fiber.  This is one of the moduli spaces
of CFT's on algebraic $K3$ surfaces discussed in \AMKthree.}
Generically, the fiber
is degenerate over 24 points on the base.  We can understand this
construction as a type IIA string vacuum as follows.  In compactifying
on $T^2$ the theory has a modulus $\tau$ describing
the complex structure of $T^2$.  In an adiabatic approximation, we can
obtain solutions to the vacuum equations by letting $\tau$ vary
slowly in spacetime (keeping the metric and the $B$-field fixed).
This can lead to interesting solutions if we
then allow $\tau$ to have $SL(2,\IZ)$ monodromy about some
submanifolds in codimension two. These solutions
were called ``stringy cosmic strings'' in
\costrings , but they differ from conventional cosmic strings in that
there is a vacuum solution even inside the core.  Since the transverse
space is four-dimensional they are more properly described as
fivebranes.  It is important to
note that the (coordinate) singularity occurs at a particular point in
the fiber (where the torus is pinched) and thus if we excise this
point the boundary of the resulting space is topologically $S^3$.
Near the core, of
course, the adiabatic approximation fails, but in at least one case
there is an exact solution reducing to it far from the core.  This is
the case in which we have 24 singularities on $S^2$,
with monodromies conjugate to $T:\tau\to\tau+1$ about each.  This can
be completed by the hyper-K\"ahler metric on $K3$ which approximates
the metric we discuss far from the singularities, and
extends across the singular fibers with no singularity in the total
space.  Of course, the product of the monodromies about the 24
singularities is the identity.

Singular points in the moduli space can be modeled in this
representation by letting the special points on $S^2$ approach each
other.  When $N$ of the singularities coincide the $K3$ manifold
acquires an $A_{N-1}$ singularity.\foot{The attentive reader will
wonder why $A_1$ singularities seem to arise in complex codimension
one, and why the famous $B$-field \refs{\gaugeKthree,\comments} has
not been mentioned.  The reason
is that we are working in a twenty-dimensional
complex slice of the full quaternionic moduli
space.} It is important to note that
the monodromies about the $N$ special points that coalesce to form
an $A_{N-1}$ singularity all commute, hence can simultaneously be
chosen as $T$.

The conformal field theory of
compactification on a torus is invariant under the exchange of $\tau$
and the K\" ahler modulus $\rho = B+i\sqrt G$, if we simultaneously
replace the type IIA theory by type IIB.  (This is mirror symmetry on
the torus.)  In the adiabatic
approximation we can thus replace a slowly-varying $\tau$ at
constant $\rho$, with slowly-varying $\rho$ at constant $\tau$.  The
singular fibers can now be taken (up to $SL(2,\IZ)$ transformations)
to be points of decompactification where the two-volume of $T^2$
diverges.  In this case the monodromy is $T: B\to B+1$ suggesting the
presence of a magnetic source of $H$-charge \OV .  Such sources are
known from a semiclassical solution, and there is a known
conformal field theory---the symmetric
fivebrane (see \chs)---which
appears to give the exact solution which is
approximated by that semiclassical one.
It has
been conjectured \refs{\comments,\OV}
that the singularity in the model with slowly-varying $\rho$
can be repaired by replacing a
small neighborhood of the singularity with the fivebrane solution.

There are two important subtleties in this case of slowly-varying $\rho$.
The first concerns the topology of the singularity.  Whereas in the
previous description the (coordinate) singularity in the theory
was concentrated at a single point, in this description in order
to obtain a smooth manifold
we must excise the entire noncompact fiber, leading to a boundary
with the topology of $S^1\times T^2$ (see \OV ).
The second subtlety in this case
is that the monodromies about different singularities on $S^2$ do
not commute.  Thus while locally near each singularity we can always
make a
choice of
$\rho$ for which the monodromy is $T$, this is
not possible globally.  This means that if we attempted to perform this
construction over all of $S^2$ at once, then about some points we would
find that the two-volume of the torus is not a single valued function of
the parameter!  Such behavior is
certainly not compatible with a description in terms of
a smooth hyper-K\"ahler manifold.
In the previous IIA description of this theory as a ``$\tau$-string'',
the local
hyper-K\"ahler manifolds near singular fibers were glued together
using transition functions which involve diffeomorphisms of $T^2$.
In the present IIB ``$\rho$-string'' description, however, the conformal
field theories must
be attached by some kind of generalized transition function which involves
a nonlocal transformation implementing the nontrivial elements in
$SL(2,\IZ)$ which act on $\rho$. Thus while one can imagine patching
together conformal field theories in an adiabatic approximation we do
not find a sigma model interpretation of the overall theory.
However, since in the semiclassical approximation it is isomorphic to
the $\tau$-string for which an exact solution is known, this unusual
construction should also lead to an exact string vacuum.

About the second point raised above we have little to add.  We will
primarily be interested in using this construction as a local model
for $N$ special fibers coalescing to the $\rho$-string analogue of
an $A_{N-1}$ singularity,
and in that context the issue is irrelevant
because the monodromies about
the singular points which coincide at the singularity all commute.

The first point is of greater concern, however, since
we would like the boundary of the singular set we remove to be $S^3$
so that the singularity can be repaired by plumbing in the
symmetric fivebrane.  This can be achieved by considering
the decompactified singular fiber as a punctured complex plane.  We
note that the $B$-fields on the fibers
 can be chosen to have compact support on $T^2$
in the vicinity of the singular fiber, approximating a
$\delta$-function as we approach the origin,
such that on the decompactified central fiber the limiting $B$-field vanishes
away from the excised origin.  (Such a construction is suggested by the
fact that the fiberwise-integrated $B$-field behaves as $\arg z$ for $z$ near 0
in the complex plane.)
In this situation we now have
a singularity in codimension four rather than codimension two.
Note that because of the choice of
$B$-field the computation of $H$-flux through $S^1\times T^2$ from
\OV\ can be deformed to find an identical flux through an $S^3$
surrounding the point singularity,
predicting the presence of a source of magnetic
$H$-charge at the point we removed.

It is natural to conjecture as in \refs{\comments,\OV} that this
source is the symmetric fivebrane.  In practice, this means that we
can cut out a small ball around the singularity and replace it with a
(suitably scaled) fivebrane solution.
The worldvolume of this fivebrane
comprises all of six-dimensional spacetime, so worldvolume dynamics
will appear as spacetime dynamics in the string theory.
Note that this description
of the singularity, and hence of the completion as well, works locally
where we have chosen $\rho$ so that the singular fiber is
decompactified (as above).
It is not clear how to describe this with a different
choice, i.e., after applying a nonlocal transformation implementing
an $SL(2,\IZ)$ shift of $\rho$.

As a final step, following \bsvi, since we are now dealing with IIB strings
we can if we wish
use $S$-duality to relate this situation to an equivalent type IIB
theory of $D$-strings, replacing the symmetric fivebranes charged under
the NS-NS field $H$ with $D$-fivebranes charged under the R-R field
$H'$.

The points of enhanced $SU(N)$ gauge symmetry are now described as
points where $N$ of these fivebranes coincide.
Because the monodromies of the coalescing singularities commute we
have a consistent manifold-like description of a neighborhood of the
singularity.
The new massless states are interpreted as open string states with
ends on two different solitons. (This is in the $D$-string picture; for NS-NS
fivebranes these are closed string states which close ``down the
throat'' \stromopen ).  The worldvolume dynamics was described in
\boundstates , and contains a dimensional reduction (to six dimensions) of
ten-dimensional supersymmetric Yang--Mills theory.
The spectrum
of light states near the singularity is thus precisely the six-dimensional
spectrum of subsection {\it 2.1}, i.e.\ we have vector multiplets of
$\CN{=}(1,1)$ supersymmetry in six dimensions with gauge group
$SU(N)$.

The picture we have reviewed above can be used
to describe the dynamics that
leads to enhanced gauge symmetry in $K3$ compactification of the IIA
string.  However, that is not the problem we wish to address here.  We
have claimed that locally about a generic point in $C$ the structure
of the transverse space is identical to the local neighborhood of an
$A_{N-1}$ singularity on $K3$.  This should allow us
to apply the description above to the $A_{N-1}$ singularity
transverse to $C$, replacing it (in a dual model of the unusual form
discussed above) with $N$ coincident fivebranes.  As in the discussion
above, the worldvolume dynamics of these is given by a six-dimensional
theory obtained by dimensional reduction of ten-dimensional super
Yang--Mills theory.  This is precisely the model studied in subsection
{\it 2.1}.  The worldvolume of
these would now be of the form $C\times \R^4$, and our problem
would be mapped to that of describing the propagation of solitons wrapped
about such a curved surface.\foot{There is a difficulty here; the
fivebranes have a tension, which would seem to imply that a $C$ of
finite size is not a solution of the equations of motion.  We thank
N. Seiberg for pointing this out to us. A more precise statement of this
problem would involve consistency criteria along
the lines of \GimPol . Such criteria are
not known for curved worldvolumes.  The twisting described above, to which we
return shortly, may resolve this difficulty. }

In order to carry out this construction, we would need something more than
a local neighborhood of $C$ in the Calabi--Yau space: we would need
what might be called a {\it semilocal}\/
neighborhood $U$ of $C$ which contains entire $T^2$ fibers (since we cannot
apply the nonlocal $\tau\leftrightarrow\rho$ transformation without
the entire $T^2$.)  Such neighborhoods will not always exist.
When they do, though,
each point in $U$ lies in a unique $T^2$ (or a degenerate limiting fiber)
so there is a well-defined map from $U$ to the parameter space $S$ for
the $T^2$ fibers (and their limits).  The curve $C$ of singularities
maps to a complex curve in $S$
which we identify with $C$, and $S$ is a small neighborhood of $C$.
Some care must be taken in interpreting this family of $T^2$'s, since
the $j$-invariant, which should be a map from $S$
to $\P1{}$, fails to be well-defined at some points $P_j$ on $C$.  (The
typical behavior is that $S$ has local complex coordinates $s_1$, $s_2$,
near such a point, and the $j$-invariant is locally given by $s_1/s_2$.)

Over $S-C$, we get a fibration by $T^2$'s and can pass from the
$\tau$-string picture to the $\rho$-string picture as above (moving to a
IIB theory).  Away from the points $P_j$, we can then complete the
$\rho$-string by including a symmetric fivebrane as discussed earlier.
However, the behavior at the points $P_j$ is somewhat mysterious.\foot{Note
that to carry out the construction as we have given it we could
actually have started with a bit less: we need a neighborhood of $C-\{
P_j\}$ rather than all of $S$.  (Such a neighborhood would be obtained from
$S$ by removing a collection of disks through the $P_j$'s, transverse to
$C$.)}  We hope
to return to this question in future work.  Assuming that the behavior at
such points is OK, the soliton indeed propagates on $C\times\R^4$ as
claimed above.

Some progress towards a description of soliton propagation in curved
spacetimes
was made in \bsvii , where it was noted that the worldvolume theory
would be twisted.  The essential point of the argument there was that
the scalars in the worldvolume supermultiplets represent transverse
motion of the $D$-brane, and thus should properly take their values in
a ``normal bundle'' which encodes the transverse motion.
When the $D$-brane wraps around a
non-flat cycle, the ``scalars'' will thus become sections
of a nontrivial bundle over this cycle.  That is precisely the situation we
have here.  The relevant scalars are the four components of $\phi$ in
\reps , which now become sections of the four-dimensional normal
bundle to $C$ encoding the transverse motion.  Two of the transverse
directions of the $D$-brane lie in the decompactified torus, and the
``normal bundle'' in those directions will be trivial.  The other two
transverse directions, though, are best described by means of the ordinary
normal bundle of $C$ within the parameter space $S$ (a complex bundle
of rank one), which can be
calculated to be\foot{The calculation is made in the IIA $\tau$-string picture,
but the result is independent of whether we are using the $\tau$-string
or the $\rho$-string model.}
\eqn\normal{
N_{C/S}=\cK_C \ ,
}
where $\cK_C$ is the canonical bundle of the complex
curve $C$.  We will verify this below.

Thus, the ``normal bundle'' to our $D$-brane worldvolume
is described as a complex bundle of rank two\foot{The bundle splits as
a $C^\infty$ direct sum, not necessarily a holomorphic one.}
\eqn\normD{
N_{D-brane} = {\cal O}\oplus{\cal K_C}\ ,
}
where $\cal O$ represents a trivial bundle of complex rank 1.
The four scalars must thus split, two remaining as scalars (a complex
section of $\cal
O$) and two one-forms (sections of $\cal K$ and its conjugate). This
uniquely determines the twist \twist .  This gives us in four
dimensions the model obtained previously, now interpreted as the
worldvolume theory.

The explicit construction we have used to find the dual description of
the singularity is, as we have seen, of limited validity (because we
require the existence of a suitable $U$).
It is not clear whether a more general argument can be made,
or whether the description in terms of parallel $D$-branes is itself
limited.  The fact that our predictions for the low-energy physics are
in agreement with those of subsection {\it 2.1\/}
suggests that the former is true, and that $U$ is only needed to
guarantee the existence of an adiabatic region.
In the next section we will verify these predictions using geometrical
methods.

\bigskip\bigskip\noindent{\it Appendix to Section 2:
Verification of Eq. \normal}\nobreak\bigskip

We resolve the singularity along $C$ with a family of ALE spaces,
obtaining exceptional divisors $E_1$, \dots, $E_{N-1}$ as described
earlier.  Let $U$ be the semilocal neighborhood which maps to the
parameter space $S$, and let $\pi:U\to S$ denote the mapping.
Then $\pi^{-1}(C)$ consists of $E_1\cup\dots\cup E_{N-1}$, together
with one more component $E_0$ which meets both $E_1$ and $E_{N-1}$.
(The fiber over a general point of $C$ consists of $n$ rational curves
arranged in a cycle, which is one form of ``degenerate limiting $T^2$''
in the classification given by Kodaira \kodaira.) If we pull back
the normal bundle of
$C$ in $S$ by the map $\pi$, we obtain the normal bundle of $E$ in
$U$, where $E=\pi^{-1}(C)=E_0+E_1+\dots+E_{N-1}$.  This can be
computed with the aid of the adjunction formula, using the fact
that the canonical bundle of $U$ is trivial.

Working with the individual components $E_i$, we find that along $E_i$,
$N_{E/U}$ takes the form
\eqn\individual{
E|_{E_i}=(E_{i-1}{\cap}E_i)+(E_i{\cap}E_{i+1})+N_{E_i/U}
}
(working with $i$ modulo $N$).  On the other hand, by the adjunction
formula (from $U$ to $E_i$)
\eqn\adjunction{
\cK_{E_i}=\cK_U|_{E_i}+N_{E_i/U}=N_{E_i/U}.
}
We can write this in the form
\eqn\kayee{
\cK_{E_i}=-(E_{i-1}{\cap}E_i)-(E_i{\cap}E_{i+1})+\cL
}
for some unknown line bundle $\cL$ on $E_i$, which must in fact be a pullback
from $C$ (since the first two terms in \kayee\ take care of the canonical
bundle of the fiber).  Since the two curves
$(E_{i\pm1}{\cap}E_i)$ are disjoint and isomorphic to $C$, if we
use adjunction from $E_i$ onto these curves we find
\eqn\adjtwo{
\cK_C=\cK_{E_{i\pm1}{\cap}E_i}=\cK_{E_i}|_{E_{i\pm1}{\cap}E_i}+
(E_{i\pm1}{\cap}E_i)^2=\cL|_{E_{i\pm1}{\cap}E_i}.
}
It follows that $\cL=\pi^*(\cK_C)$.

Since this happens on each $E_i$, we see that $N_{E/U}=\pi^*(\cK_C)$,
and so $N_{C/S}=\cK_C$, verifying \normal.

\newsec{Geometry of the Moduli Space}

In this section, we describe the geometry of the moduli space, find the
extremal transition corresponding to the Higgs branch, and verify the
predictions we have made concerning the structure of the moduli space,
the monodromy of the periods, and the topological one-loop amplitude $F_1$.

\subsec{The Vector Multiplet Moduli Space}

To compute the vector multiplet moduli space, we represent our singularity
on a Calabi--Yau hypersurface $M$ in a toric variety $V$.\foot{Both the
geometry and the physics that we are describing will only depend on a
small neighborhood of the singular curve; the use of toric geometry is
a technical convenience for analyzing this neighborhood.
For a review of toric geometry, see \agm.}
We follow the
description given by Batyrev \rbat\ of such hypersurfaces in terms of
a {\it reflexive polyhedron}\/ $\D$, which specifies the monomials occurring
in the equation of $M$.
The polar polyhedron
$\D^\circ$ is also reflexive.  The Calabi--Yau hypersurface will have
a singularity as described above precisely when
 $\D^\circ$ has an edge
joining two vertices $v_0,\ v_N$ of $\D^\circ$ with $N-1$ equally spaced
lattice points $v_1,\ldots,v_{N-1}$ in the interior of the edge.  It is
assumed that $N\ge 2$.  The Calabi--Yau threefold $M$ is a hypersurface
in the toric variety whose fan is a suitable refinement of the fan
consisting
of the cones over the faces of $\D^\circ$.  In particular, there are
edges corresponding to the lattice points $v_1,\ldots,v_{N-1}$.  These
correspond to toric divisors, which resolve a surface $S$ of $A_{N-1}$
singularities in $V$.  Restricting to the hypersurface $M$, we see that
there are $N{-}1$ divisors in $M$ which resolve a curve $C$ of $A_{N-1}$
singularities.

The genus $g$ of $C$ can be easily determined by toric geometry.  By duality,
the edge $D_1^*=\langle v_0,v_N \rangle$ determines a two-dimensional
face $\D_2$ of
$\D$.  The number of integral interior points of $\D_2$ is equal to the
genus $g$.

The mirror manifold $W$ is constructed \rbat\ as a hypersurface in a toric
variety $\varLambda$.  Its defining equation is
a sum of monomials corresponding to the integral points of
$\Delta^\circ$.  By our assumptions, there will be an edge with
integral interior points.  The linear equations
\eqn\linear{2v_k=v_{k-1}+v_{k+1} }
for $k=1,\ldots,N-1$ translate into relations on the
monomials $m_i$ associated to these $v_i$,
\eqn\emrel{
m_{k-1}m_{k+1}=m_k^2\ .
}
If the coefficient of $m_i$ in the defining equation is denoted $a_i$
then \emrel\ shows that $W$ is singular on the components of the
divisor in moduli space given by the vanishing of the
discriminant of the degree $N$ polynomial
\eqn\Npoly{
a_0x^N+a_1x^{N-1}y+\ldots+a_Ny^N\
}
in two auxiliary variables $(x,y)$.

\subsec{The Type IIB Description}

This structure leads to a subtlety in the description of the moduli
space of the conformal field theory near the singularity.  At a smooth
point the moduli space factors as a product of the space of complex
structure deformations and the space of (complexified) K\"ahler
deformations.  Near the singularity this factorization fails,\foot{This was
independently noted in \bsvi.} for the
following reason.  There are toric divisors
$V_i$ in $\varLambda$ associated to the $g$ interior points of the face
$\Delta_2$, and when restricted to $V_i$ the equation for $W$ becomes
{\it precisely\/} $a_0m_0+a_1m_1+\ldots+a_Nm_N$. (One can think of
this heuristically as a result of considering the $V_i$ as exceptional
divisors resolving a quotient singularity in the ambient space which
coincides with the singularity of $W$.)
Generically the restriction has $N$ components.  As usual, this leads to
$g$ sets of $N{-}1$ non-toric divisor classes on $W$.  Along the discriminant
divisor discussed above, these coalesce; at a generic point
one component acquires multiplicity $2$ along each $V_i$.  The divisor
classes thus
undergo $S_N$ monodromy about the components of the singular divisor.
Because the ``sum'' of each $N$ components is the toric divisor, the
non-toric classes transform naturally in the $(N{-}1)$-dimensional
representation.

Thus, the discriminant locus has a natural stratification
indexed by partitions of $\{1,\ldots,N\}$ into subsets describing the
coalescing of the roots of \Npoly.  The minimal stratum occurs where
\Npoly\ is a perfect $N^{\scriptstyle\rm th}$ power of a linear monomial
which we will denote as $z$.  The $g$ interior integral points of
$\Delta_2$ give some of the homogeneous coordinates of the toric variety.
We will denote these as $x_1,\ldots x_g$.
The hypersurface $W$ is defined by an linear combination of monomials
corresponding to integral points of $\Delta^\circ$.  The points of
$D_1^*$ correspond to the only monomials which do not involve the variables
$x_1\ldots,x_g$.  At the minimal stratum, the combination of these
monomials is precisely $z^N$.  Thus the equation of $W$ is of the
form
\eqn\Veqn{
z^N=x_1\cdots x_g f
}
where $f$ depends on the choice of complex structure.  Putting
$y_i=x_i$ for $1\le i\le g$ and $y_{g+1}=f$, we see from \Veqn\ that the
singular locus of $W$ is the union of the curves $z=y_i=y_j=0$ for
each pair of indices $(i,j)$ with $i\neq j$.\foot{In \twoparamsI, the example
$\P4{(1,1,2,2,2)}[8]$ was considered, and the singularity we are discussing
here was described there as being isolated.  However, the model used in
\twoparamsI\ must be blown up further in order to be able to realize generic
K\"ahler structures, and we again reach the description just given of the
curve of singularities.} It is important that
\Veqn\ is written in homogeneous coordinates on $\varLambda$, subject
to the restrictions on intersections of coordinate hyperplanes that
follow from the toric structure.

We can also write down a local equation for the general
deformation.  It is clear that this is done by deforming the left-hand
side of \Veqn\ to a generic polynomial of degree $N$ in $z$.  The
coefficients of this are some functions $x$ such that the deformed
equation is still quasihomogeneous.  Choosing a representative $f_j$
in each relevant degree we can write the deformation as
\eqn\Vdefeqn{
z^N + \sum_{j=0}^{N-2} u_j f_j = x_1\cdots x_g f
\ .}
The coefficients $u_j$ give local coordinates on the deformation
space.  The moduli space of K\" ahler structures is a nontrivial $S_N$
bundle over this, as noted above.  To obtain a product structure we
pass to an $S_N$ cover, introducing coordinates
$(z_1,\ldots,z_N)$ satisfying the constraint $\sum_iz_i=0$, such that
the $u_j$ is the symmetric polynomials of degree $j$ in $z_i$ (the
constraint simply reflects $u_{N-1}=0$).
With the same notation as above, we write the equation as
\eqn\genVeqn{
\prod_{i=1}^N (z-z_ih_i) =x_1\cdots x_g f
}
for some other functions $h_i$.
The $z_i$ are the direct analogs of the $\phi_i$ of the previous
section.

To reiterate: once we have passed to an appropriate $S_N$ cover of the
complex structure moduli space, the moduli space of conformal field
theories splits as a product of the complex structure moduli and the
K\"ahler moduli.  The group $S_N$ acts on this, with nontrivial action
on both factors in the product, and the moduli space of CFTs is the quotient
of the product by this group action.

One signal of this subtle structure is the behavior of the periods of
the holomorphic $3$-form.  (In fact, this is how we
originally discovered this phenomenon.\foot{A similar discovery---of a
double cover of the moduli space in an example with $N=2$---was made
independently in \ghl, based upon an analysis of the heterotic
weak-coupling
enhanced gauge symmetry point.})  The periods which are related
to coordinates on this subspace of the moduli space can be calculated
with the aid of the generalized hypergeometric differential equations
of Gel'fand, Zelevinski\u\i\ and Kapranov \GZK.
These periods have a monodromy near the discriminant locus which reproduces
the $S_N$ cover.
The combinatorics governing the
differential equations are precisely given by \linear; in the case
of $N=2$ these equations
were solved in \small.  The general solution can be written as\foot{We
are using a coordinate $t=a_0a_2/a_1^2$
which differs from the one used in \small\ by a factor of $4$.}
\eqn\gensol{C_1+C_2\log\left({1-2t-\sqrt{1-4t}\over2t}\right) }
(with $C_1$ and $C_2$ being constants of integration),
although the structure we are interested in is more visible if we
exploit the identity
\eqn\exploitid{
\log\left({1-2t+\sqrt{1-4t}\over t}\right)
+ \log\left({1-2t-\sqrt{1-4t}\over t}\right) = \log4
}
to rewrite \gensol\ as
\eqn\gensolbis{C'_1\log\left({1-2t+\sqrt{1-4t}\over t}\right)
+ C'_2\log\left({1-2t-\sqrt{1-4t}\over t}\right) \ .}
The discriminant is located at $t=1/4$, and if we analytically continue
around that point,  the $S_2$ action is manifest.

The generalization to arbitrary $N$ can be done with the methods of \small,
but it is simpler to see the monodromy of the periods from the IIA
perspective, to which we turn next.

\subsec{The Type IIA Description}

There is an analogous subtlety using the mirror description
of the theory via the IIA string on $M$.  This is automatic due to
mirror symmetry, but it will prove useful to go through this in some
detail.

Here the vector moduli space consists of the
complexified K\"ahler classes of $M$.  To see the non-product
structure, though, we need to enlarge the K\"ahler moduli space.
(We know from mirror symmetry that such enlargements are often
necessary \agm.)

Let us consider a general situation in which
the K\"ahler cone $\cK_M\subset H^2(M)$ has a codimension one
face of the form $\cK_M\cap\Gamma^\perp$,
where $\Gamma\in H^2(M)$ is the class of a holomorphic $2$-sphere.
The topological $3$-point functions for the conformal field theory
take the general form
\eqn\threepoint{
\langle\cO_A\cO_B\cO_C\rangle=A\cdot B\cdot C+\sum_{\eta\in H_2(M)}
{q^\eta\over1-q^\eta}\,N_\eta\,(A\cdot\eta)(B\cdot\eta)(C\cdot\eta)
}
for $A, B, C\in H^2(M)$, where $N_\eta$ counts the number of holomorphic
$2$-spheres of class $\eta$ (in the generalized sense of ``Gromov--Witten
invariants''), where $q^\eta$ denotes $\prod q_i^{(E_i\cdot\eta)}$ once a basis
$\{E_i\}$ for $H^2(M)$ has been chosen, and where we have incorporated
the ``multiple cover formula'' of \refs{\cdgp,\tftrc}.
The analytic continuation of this expression to the
other side of
the wall $\Gamma^\perp$ can be computed using the identity
\eqn\identity{
{q^\Gamma\over1-q^\Gamma}=-1-{q^{-\Gamma}\over1-q^{-\Gamma}},
}
which results in an expression of the form\foot{Note that the location
of the singularity in the three-point functions asymptotically approaches
$q^\Gamma=1$ when we tune the other parameters so as to suppress all other
instanton contributions \small.}
\eqn\threepointbis{\eqalign{
\langle\cO_A\cO_B\cO_C\rangle&=\left(A\cdot B\cdot
C-N_\Gamma\,(A\cdot\Gamma)
(B\cdot\Gamma)(C\cdot\Gamma)\right)\cr
&+{q^{-\Gamma}\over1-q^{-\Gamma}}\,N_\Gamma
\,(A\cdot(-\Gamma))
(B\cdot(-\Gamma))(C\cdot(-\Gamma))\cr
&+\sum_{\eta\in H_2(M), \eta\ne\Gamma}
{q^\eta\over1-q^\eta}\,N_\eta\,(A\cdot\eta)(B\cdot\eta)(C\cdot\eta)\
.\cr
}}

Suppose that we have another Calabi--Yau manifold $\widehat M$ and
an isomorphism $\rho:H^2(M)\to H^2(\widehat M)$ such that
the instanton numbers $N_\eta$ are preserved, and such that the
cones
$\cK_M$ and $\rho^{-1}(\cK_{\widehat M})$ meet along their common
wall $\Gamma^\perp$.
If
\eqn\abc{
\rho(A)\cdot \rho(B)\cdot \rho(C)=A\cdot B\cdot
C-N_\Gamma\,(A\cdot\Gamma)
(B\cdot\Gamma)(C\cdot\Gamma),
}
then \threepointbis\ shows that
the sigma model on $\widehat M$ provides an analytic continuation
of the sigma model on $M$ across the wall $\Gamma^\perp$.

Perhaps the most familiar instance of this construction is the case
of a flop, where this was worked out in \refs{\phases,\small} (see
also \beyond\ for more details).  If there are a finite number $N_\Gamma$
of curves in the class $\Gamma$, then flopping them all produces
the Calabi--Yau manifold $\widehat M$ and the map $\rho$ is the ``proper
transform'' of divisors.  (Eq.~\abc\ was verified for this case in \beyond.)
{}From this calculation we learn that the cones $\cK_M$ and
$\rho^{-1}(\cK_{\widehat M})$
can be attached along their common face to form a moduli space of
a single class of conformal field theories with
two different phase regions (corresponding to sigma models on
different Calabi--Yau
manifolds).  There is an obvious extension of this to the union of
the K\"ahler cones of all possible flops of $M$ (transported back to
$H^2(M)$ as $\rho^{-1}(\cK_{\widehat M})$) which forms the
{\it partially enlarged K\"ahler moduli space}\/ of \agm.  (The
corresponding cone is known as the {\it movable cone}.)

A less familiar instance of the same construction
 is the case of an {\it elementary
transformation}.\foot{Elementary transformations were first discussed
for $K3$ surfaces in
\BR\ and for Calabi--Yau threefolds in \Wilson\ in the case of $g=1$;
elementary transformations over curves of arbitrary genus were
discussed in section 9 of \twoparamsI,
and section 6 of \beyond, but not in much detail in either place.}
Such a transformation exists whenever the class $\Gamma$ is represented by a
family of holomorphic $2$-spheres which sweep out a divisor $E$.
(This divisor is contracted to a curve of genus $g$ when the
area of the $2$-spheres goes to zero.)
To define the elementary transformation we take $\widehat M=M$
and define the isomorphism $\rho$ by
\eqn\reflection{\rho(H)=H+(H\cdot\Gamma)E.}
Since $E\cdot\Gamma=-2$, this is a {\it reflection}, i.e., $\rho(E)=-E$.

To verify eq.~\abc\ for an elementary transformation requires a bit of work.
Since $E^2=K_E$ by the adjunction formula, we have
$E^3=(K_E\cdot K_E)_E=8-8g$.  Furthermore, denoting by $\Gamma$ a divisor
class on $M$ whose restriction to $E$ is $\Gamma$, we have
$(A|_E\cdot\Gamma)_E=A\cdot\Gamma$ and
$(E|_E\cdot\Gamma)_E=(K_E\cdot\Gamma)_E=-2$, so the combination
${1\over(A\cdot\Gamma)}A\cdot E+{1\over2}E^2$ is a cycle on $M$ which
lies on $E$ and
which has intersection number $0$ with $\Gamma$; it is
therefore numerically equivalent
to $a\Gamma$ for some $a$.  Similarly, there are numerical
equivalences
\eqn\numequiv{
{1\over(B\cdot\Gamma)}B\cdot E+{1\over2}E^2\equiv b\Gamma, \qquad
{1\over(C\cdot\Gamma)}C\cdot E+{1\over2}E^2\equiv c\Gamma.
}

We now compute
\eqn\bigcomp{
\eqalign{
{1\over(A\cdot\Gamma)}A\cdot E^2&
=
-{1\over2}E^3 +a (E\cdot \Gamma) = (4g-4)-2a;\cr
{1\over(A\cdot\Gamma)(B\cdot\Gamma)}A\cdot B\cdot E
&=
-{1\over2(B\cdot\Gamma)}B\cdot E^2
+{a\over(B\cdot\Gamma)}(B\cdot\Gamma)
=
(2-2g)+b+a.\cr}
}
It follows that
\eqn\concl{
\eqalign{
&
{{\rho( A)}\cdot {\rho( B)}\cdot {\rho( C)}-A\cdot B\cdot C\over
(A\cdot\Gamma)(B\cdot\Gamma)(C\cdot\Gamma)}
\cr
&\quad\qquad\qquad=
{(A+(A\cdot\Gamma)E)\cdot (B+(B\cdot\Gamma)E)\cdot
(C+(C\cdot\Gamma)E)-A\cdot B\cdot C\over
(A\cdot\Gamma)(B\cdot\Gamma)(C\cdot\Gamma)}\cr
&\quad\qquad\qquad=(2-2g+b+a)+(2-2g+c+a)+(2-2g+c+b)\cr
&\quad\qquad\qquad\quad+(4g-4-2a)
+(4g-4-2b)+(4g-4-2c)+(8-8g) \cr
&\quad\qquad\qquad=(2-2g)\cr }
}
This agrees with~\abc\ since we know that
$N_\Gamma=2g-2$ as explained in~\twoparamsI\ (see also the next
subsection).

The interpretation of the elementary transformation is a bit different
than that of the flop.  The analytic continuation of the conformal field
theory from the cone $\cK_M$ to the cone
$\rho^{-1}(\cK_M)$ produces a theory which, by virtue of the
elementary transformation $\rho$, is isomorphic to the sigma model on $M$,
but which uses a different basis for $H^2(M)$.  We can enlarge the
moduli space further
to include these reflected cones $\rho^{-1}(\cK_M)$,
but since we are obtaining identical theories we should then mod out by
the action of the reflection.  This would appear to be a trivial operation
were it not for the fact that when we vary the complex structure, the
cones involved can change and the elementary transformation can cease to exist.

More precisely, what happens is this:  if $g{\ge}1$, the complex structure
can be deformed so that the class $\Gamma$ is represented by only finitely
many holomorphic 2-spheres ($2g{-}2$ of them, to be precise).  In the case
of $g{=}1$, the K\"ahler cone also changes: the generic K\"ahler cone
is the union of the cones which appeared at special complex structure.
In the case of $g{>}1$ it is the movable cone which changes: the various
cones are still glued together, but at generic complex structure they
represent Calabi--Yau manifolds related to each other by flopping.

The flops, in this case of a perturbed elementary transformation, actually
take the Calabi--Yau manifold to another Calabi--Yau manifold of the
same type, {\it but with a different complex structure and different
K\"ahler class}.  Thus, the
elementary transformation $\rho$ actually acts on the entire
moduli space, moving from point to point, identifying Calabi--Yau's which
differ by a flop.  (These identifications change both the complex
and the K\"ahler structures simultaneously.)
The transformation $\rho$ fixes a subset of the complex structures---the ones
for which the divisor $E$ exists---and on that fixed locus it acts as
the elementary transformation described above.

It is now clear how $S_N$ acts in this situation, when there is
a curve of $A_{N-1}$ singularities.
We first observe that the $N{-}1$
elementary transformations associated to the $N{-}1$ exceptional
divisors generate the group $S_N$.  This gives an $S_N$ action on the
complex moduli space. On the part of  the moduli space which is
transverse to the locus where the $E_i$'s exist,
we expect this representation to be a sum of
$g$ distinct $(N{-}1)$-dimensional representations according to our
discussion of the IIB string on $W$.  This assertion will be verified
directly later using deformation theory.

The $S_N$ action on the complex structure moduli space must be coupled to its
action on the enlarged K\"ahler moduli space.  In fact,
 the ``reflected movable cone'' \beyond, which is the union of all
translates of the
K\"ahler cone under all possible flops and
reflections, is now invariant under deformation of complex structure,
and is acted upon by the $S_N$ action according to \reflection.
This is our other $(N{-}1)$-dimensional
representation of $S_N$.  Our descriptions now agree for the
IIA and IIB models: in each case, we have a product of complex
structure and K\"ahler moduli, with an action of $S_N$ on each factor,
and we take the quotient of that product by the combined action
to obtain the correct CFT moduli space.

\subsec{Geometry of the Extremal Transitions}

In the remainder of
this section, we consider the extremal transitions and show several
ways in which the transitions are in perfect agreement with the
gauge theoretic predictions.
First, let us quickly clarify our meaning of an extremal transition,
following \trento.
This class of transitions generalizes the notion of a conifold transition.
We start with a Calabi-Yau threefold $M$ which admits a birational
contraction $M\to N$ to a singular threefold $N$.  If $N$ can be deformed to a
smooth Calabi-Yau threefold $N'$, then the transition
from $M$ to $N'$ is called an extremal transition.  Recent progress in
understanding when $N$ is smoothable has been made in
\refs{\NS\grossdef{--}\rgross}.

Our first prediction from the gauge theory is the existence of a Higgs
branch, which should correspond to an extremal transition from $M$
to another Calabi--Yau manifold $N'$ whose Hodge numbers are such as
to produce a moduli space with dimension in accordance with \ehigs.
It is easiest to see the transition geometrically in the IIA description.
Here, we reach the transition
by contracting all exceptional divisors to our curve $C$ of $A_{N-1}$
singularities (with an appropriate value of the $B$-field).  It was shown
recently that this singular threefold deforms to a smooth Calabi--Yau
threefold~\rgross.  This is the desired transition.
It would be very desirable to be able to see the transition in the IIB
description.

To compute the change in Hodge numbers associated to this transition
(for comparison to \ehigs ) it proves most efficient to proceed along
the alternate path mentioned in subsection {\it 2.2}\/
which avoids the point of
enhanced gauge symmetry.
In effect, these examples now give us conifold transitions in
type IIA string theory, and this will allow us to calculate the
dimensions of the moduli spaces and show that they agree with our
gauge theoretic predictions.  From the point of view of the gauge
theory these transitions are identical to those obtained in \gms , but
we will describe them here in the language appropriate to type IIA
strings, whereas the discussion in \gms\ was in terms of type IIB
string theory.\foot{The transitions are identical provided that
the mirror of a conifold transition is a conifold transition---something
which was implicitly conjectured in \refs{\trento,\gms}.}

We claim that after we effect a general non-toric
deformation, the exceptional divisors $E_i$ are replaced with
$N\choose 2$ collections of $2g{-}2$ homologous 2~spheres.
These 2~spheres can then be used to give a conifold transition.

In a little more generality,
suppose that we have a Calabi--Yau threefold obtained by
resolving a genus $g$ curve of $A_{N-1}$ singularities in a singular
Calabi--Yau
threefold.  The exceptional divisors
form a chain of $N{-}1$ irreducible ruled surfaces over this curve
of genus $g$, which we order consecutively so that $E_i$ meets $E_{i+1}$.
In particular,
this is the situation for the manifolds $M$ that we are considering.

For any subset $\{j,j+1,\ldots,k-1,k\}$ of $\{1,\ldots,N-1\}$, we consider
the fibers of $E_j\cup\cdots\cup E_k$.  This is a family of chains of $\P1{}$s
parameterized by the singular curve $C$.  Its deformation theory is controlled
by its normal bundle since it is still a local complete intersection in the
threefold.  The tangent space to the deformation space is $H^0(\Normal)$, which
is then identified with the tangent space to $C$.  The obstruction space
is $H^1(\Normal)$.  The argument of \twoparamsI\ applies.  We have a perfect
pairing
\eqn\perfpair{
H^0(\Normal)\otimes H^1(\Normal) \to H^1(\wedge^2 \Normal) \simeq
H^1(\omega),
}
where $\omega$ is the canonical sheaf of our reducible curve.
Note that $H^1(\omega)$ is one-dimensional.  Furthermore, by duality, this
pairing globalizes to a pairing of line bundles over $C$.  The right
side of the pairing globalizes to the trivial bundle.  The space
$H^1(\Normal)$ globalizes to the obstruction bundle.  The pairing now shows
that
the obstruction bundle is the cotangent bundle of $C$.

Returning for definiteness to our class of examples, i.e., the ones
realized as toric hypersurfaces, it has been remarked
earlier that there is a $g(N{-}1)$-dimensional space of non-toric
deformations \rbat.
A general deformation induces a section of the
obstruction bundle, which has $2g{-}2$ zeros corresponding to the fibers
which will deform.  In other words, a general (non-toric) deformation
contains $N\choose 2$ collections of $2g{-}2$ rational curves, one
for each choice of $j,k$.  This completes the justification of our
claim.\foot{This was the argument used in \twoparamsI\ to establish that
$N_{\Gamma}=2g-2$.}
This deformation should correspond to the region of moduli space
discussed in subsection {\it 2.1}\/ following equation \emix , in which
we make the transition to the Higgs branch avoiding the point of
enhanced gauge symmetry.  At first sight there is an
apparent discrepancy in the two descriptions. In the gauge theory, we
predict the existence of $g{-}1$ hypermultiplets with the charges
under $U(1)^{N-1}$ of the adjoint representation.  Here we find
instead twice as many cycles with the charges of the positive roots
(as mentioned, the neutral components are not solitonic in origin and
are manifest).  Of course, the discrepancy is only in appearance,
because the adjoint representation is real.  Thus considering a double
set of hypermultiplets with only the positive charges is simply a
reshuffling of the fields.

It is illuminating to consider this in more detail.
The $g(N{-}1)$ complex structure deformations can be seen explicitly.
For each of the exceptional divisors $E_i$, there is a map
$H_1(C)\to H_3(M)$ given by sending a 1~cycle $\gamma$ on $C$ to the 3~cycle
swept out by the fibers of $E_i$ lying over $\gamma$ \rCG.  This
can be refined to a mapping $\phi_i:H^{1,0}(C)\to H^{2,1}(M)$.  The images
of the $N{-}1$ maps $\phi_i$ span the non-toric deformations.  This
geometrical discussion precisely reflects the dimensional reduction
following \twist ; the hypermultiplets (complex structure
deformations) are related to the same two-cycle $\Gamma_i$ as the
corresponding vector multiplets (the modulus dual to $E_i$), but are
related to one of the $g$ holomorphic one-forms on $C$.

Similarly, there are maps $\phi_{i,j}:H^{1,0}(C)\to H^{2,1}(M)$ defined
using the reducible fibers of $E_i\cup\ldots\cup E_j$.  The set of all
such unions $E_i\cup\ldots\cup E_j$ for $i\le j$ are naturally identified
with the set of positive roots of the root system $A_{N-1}$ by identifying
it with the root $e_i-e_{j+1}$ in the notation of \rBour.
With this identification, the
Weyl group of $A_{N-1}$ is $S_N$ and
naturally acts on the space of non-toric deformations
by exchanging the images of the $\phi_{i,j}$ as dictated by the
corresponding
Weyl group action on the abstract root system.  These classes are not
independent of those described in the previous paragraph, they give
another (overcomplete) basis for the same space, which makes the $S_N$
action manifest.  It is clear from this
description that we have $g$ distinct $(N{-}1)$-dimensional representations,
as was asserted in our earlier discussion.  The discriminant locus in
this context will occur for certain K\"ahler structures which contract
one of the $E_i$ (with a suitable shift by the $B$ field), since
some of the fibers in the above description will now coalesce.  This again
agrees with what we found using the IIB string on $W$.

As an alternative approach to understanding the structure of the non-toric
deformations of $M$, we can start by
considering  deformations where the exceptional divisors
deform to $2g{-}2$ chains of $N{-}1$ connected $\P1{}$s.  The general
case can be inferred from the deformation theory of these
chains.  Criteria for when a connected
union of $\P1{}$s can be deformed or contracted are
considered in the doctoral thesis of T.~Zerger \zerger.

There is a part of the boundary in the K\"ahler moduli space of $M$ on which
the area of each of the fibers of the $E_i$ approach 0.  Our transition
will occur somewhere on here by supplying an appropriate value of the
$B$-field.
Remaining at this location in the K\"ahler moduli space and performing
a general complex structure deformation as discussed above, we
are in a situation where $(2g{-}2){N\choose 2}$ curves get contracted to
nodes.  These curves span a $(N{-}1)$-dimensional subspace of the
homology of $M$  (generated by the classes of the $N{-}1$ irreducible
fibers).

Now smoothing these singularities, we claim that the Hodge numbers change
by
\eqn\Hodgechange{
h^{1,1}\mapsto h^{1,1}-(N-1),\quad h^{2,1}\mapsto
h^{2,1}+(2g-2){N\choose 2}-(N-1).
}

To see this, note first that the
drop of $N{-}1$ in $h^{1,1}$ is clear---the transition kills $N{-}1$
independent homology 2-cycles.
Furthermore, the transition is obtained homotopically by replacing
$(2g{-}2){N\choose 2}$ two spheres by $(2g{-}2){N\choose 2}$ three spheres.
This results in a change of Euler characteristic by
$-2(2g{-}2){N\choose 2}$.
Since we
have already accounted for $-2(N{-}1)$ two-dimensional and four-dimensional
homology, there must be a net
change of $2(2g{-}2){N\choose 2}-2(N{-}1)$ in three-dimensional homology.
These are shared equally by $H^{2,1}$ and
$H^{1,2}$, giving the claimed result.

Note that the above argument more generally applies to describe the
conifold transition in type IIA string theory.
Suppose that $R$ rational curves spanning an $S$-dimensional subspace of
homology are contracted and then smoothed. We have
\eqn\dimchange{
h^{1,1}\mapsto h^{1,1}-S,\quad h^{2,1}\mapsto h^{2,1}+R-S
}
for the change in Hodge numbers.  In this case we have a clear
understanding, following \refs{\andy,\gms} of the solitonic states
responsible for the new massless particles.
These are obtained by wrapping
2~branes (presumably $D$-branes of the type IIA string) around
2~spheres.  These are charged with respect to the R-R
$U(1)$'s corresponding to the K\"ahler moduli; but by our assumptions
the charges factor through a $U(1)^S$.
These states become massless at the transition, and the
transition occurs by Higgsing the $U(1)^S$.  Examples of these transitions
are given in \refs{\rbkkI,\rbkkII}; in the latter example, it is shown that
the gauge-theoretic description of the transition we give here matches with
the description of the transition between the heterotic duals.

\subsec{Monodromy}

Our gauge-theoretic model has given us a prediction for integral
monodromy.  We verify this using the IIA interpretation of the
model.  Rather than dealing directly with period integrals, we calculate
with their mirrors.

Our verification is based on the connection between period integrals
and Yukawa couplings dictated by special geometry \rStrom.  On the
type IIB side, if we choose a holomorphic three-form $\Omega$
and a symplectic basis $\{\alpha_i,\beta^i\}$
for the homology cycles, then the corresponding periods
\eqn\corrper{
a_i:=\int_{\alpha_i}\Omega \ , \qquad a_D^i:=\int_{\beta^i}\Omega
}
are related to the prepotential $\cF$ through the formula
\eqn\defaD{a_D^i={\partial \cF\over \partial a_i} \ .}
On the other hand, the cycles $\alpha_i$ determine marginal operators
$\cO_i$, the three-point functions among which are
\eqn\Yuk{\langle\cO_i\cO_j\cO_k\rangle={\partial^3\cF\over
\partial a_i\partial a_j\partial a_k} \ .}
We can thus determine the behavior of the periods from the behavior of
the three-point functions.  The $a_i$'s
are the flat coordinates in this space; on the mirror model, these
are the coefficients in an expansion of the K\"ahler class with
respect to a fixed topological basis of $H^2(M)$.

On the type IIA side, we have found an explicit description of the
moduli space and the three-point functions.
We only need to calculate the monodromy associated with
one of the reflections which generate $S_N$, which we have previously
represented as an elementary transformation.
The 4-cycle $E$ and
the 2-cycle $\Gamma$ of this transformation
form a two-dimensional subspace of the even cohomology
which contains everything which is varying as we perform our monodromy.
Note that we can work on the covering space of the moduli space which
is a product of K\"ahler and complex moduli, where the path along
which we calculate monodromy is not closed.

We choose a basis of $H^2(M)$ which includes $E$ as the first basis element,
and let $a_1$ be its coefficient in the expansion of the K\"ahler class.
Our singularity lies at $q_1=1$.
{}From \threepoint\ we see that the three-point function
$\langle \cO_E \cO_E \cO_E \rangle$ has a simple pole there with leading term
\eqn\leading{
{q_1^{-2}\over 1-q_1^{-2}} N_\Gamma (E\cdot\Gamma)^3\ ;}
 moreover, we know that $N_\Gamma=2g-2$ and
$E\cdot\Gamma=-2$.  The area of the 2-sphere $\Gamma$
is $a=-2a_1$. Our path connects a base point to its image under
the elementary transformation $\rho$; since $\rho$ maps $E$ to $-E$,
the map on the coefficient must be $a_1\mapsto -a_1$.  This gives the
action of the monodromy on $a$ as being $a\mapsto -a$.

To compute the monodromy action on $a_D$ we must be a bit careful.
Although
the classes $E$ and $\Gamma$ span the generalized eigenspace with
eigenvalue
$-1$ for the monodromy transformation, they are not part of a symplectic
basis for $H^{even}(M,\IZ)$.  (They are the closest we can come to a
symplectic, integral basis within that space however.)  Since the
partial derivative $\partial\cF/\partial a$ gives the period corresponding
to the symplectic partner of $E$, i.e., corresponding to $-\Gamma/2$,
the period $a_D$ corresponding to $\Gamma$ must be given by
$-2\,\partial\cF/\partial a$.

To find the monodromy action, we
calculate using \Yuk\ and \leading:
\eqn\thecalc{
{\partial^3\cF\over\partial a^3} = -{1\over8}{\partial^3\cF\over\partial a_1^3}
=-{1\over8}\langle \cO_E \cO_E \cO_E \rangle
= {q_1^{-2}\over 1-q_1^{-2}} (2g-2) +\cdots
={-1\over2\pi ia}(2g-2) +\cdots \ .
}
Integrating twice and multiplying by $-2$, we obtain
\eqn\theanswer{
a_D\sim {2g-2\over\pi i}a \log a \ ,}
verifying \eas\ and hence \emon.
In particular, we see here the logarithmic corrections to $a_D$, the
monopole mass, arising from world-sheet instanton effects.

It is also possible---at least in a few examples---to
directly verify \emon\ for the period integrals in
a IIB interpretation of the model.  (These calculations are expected
to be equivalent in general according to the conjectures of \predictions.)
We will discuss this direct verification in the next section.

\subsec{One-Loop Amplitude}

We can calculate the leading power-law behavior of the one-loop
topological amplitude
$F_1$ for all of our examples.  This can be done
following methods developed in \refs{\BCOV\HKTYI{--}\HKTYII,\rbkkI}.

We consider the IIA theory on $M$, and start by calculating the
K\"ahler cone of $M$ using either toric geometry or the calculation of the
secondary fan.\foot{It should be noted that the K\"ahler cone of the
toric variety can be different from the K\"ahler cone of $M$
\refs{\mondiv,\rbkkI}.}
In the sequel, we will denote the primitive integral generators
of the K\"ahler cone by $J_1, J_2,\ldots, J_k$.
The Mori cone of $M$ can then be computed as the dual of its K\"ahler
cone.  This gives coordinates $z_i$ on the instanton corrected
K\"ahler moduli space of $M$, or equivalently on the complex structure
moduli space of $W$.

Using the methods of \refs{\HKTYI,\HKTYII,\rbkkI} we can calculate the
principal parts of the Picard--Fuchs differential operators, and input
this data and the Mori cone into the program {\sc instanton} which was
appended to the hep-th version of \HKTYII.  This gives us the
mirror map (the flat coordinates $t_i$ as a function of the algebraic
coordinates $z_i$) and the fundamental period $\varpi_0$.

{}From the generalization of the formula of \BCOV\ to Calabi--Yau
threefolds with more than one modulus we have
\eqn\Ftop{F_1^{\rm top}=\log\left({1\over\varpi_0^{3+h^{1,1}(M)
-{\chi(M)\over 12}}}f(z) {\partial (z_1,\ldots,z_k) \over
\partial (t_1,\ldots,t_k)}\right)}
for some function $f(z)$ which is holomorphic on the interior of the
moduli space.  The boundary of the moduli space consists of two types
of divisors---toric
boundary divisors and components of the discriminant locus.
The toric boundary divisors are defined in our coordinates
by monomials in the $z_i$.  The components of the discriminant will
be denoted by $\Delta_i$.  With these conventions, the unknown
holomorphic function $f$ above must take the form
\eqn\unknown{
f=\prod_i \Delta_i^{r_i} \prod_j z_j^{s_j}
}
for some unknown exponents $r_i,\ s_j$.  The calculation of $F_1^{\rm top}$
is thus reduced to the calculation of these exponents.
We order the components $\Delta_i$
so that $\Delta_1$ is the principal component of the discriminant and
$\Delta_2$ is the component along which our transition occurs, namely the one
which we have identified with the discriminant of~\Npoly.

The calculation proceeds in two steps.  We first use the asymptotic
result from \BCOV\ (where the normalization of the $t_i$ differs from
ours by a factor of $2\pi i$)
\eqn\asympt{
F_1^{\rm top} \sim -{2\pi i\over 12}\sum_i {c_2(M)\cdot J_i \over 12} t_i
}
to express the $s_j$ in terms of the $r_i$.  Next, the numbers of elliptic
curves on $M$
can be computed from the expansion of
${\partial\over \partial t_i}F_1^{\rm top}$ in terms of the $s_i$.

The next step is to use simple geometric arguments to show that
there are no elliptic curves in certain low degree homology classes.
This enables us to solve for the $r_i$.  We will illustrate the procedure
in an explicit example in the next section.

In our cases, we will always have a component of the discriminant
corresponding to the transitions we are considering here.  In all
cases, the corresponding exponent $r_i$ agrees with the results
predicted from our gauge-theoretic description.  To see this, note
that  the prediction \efi\ in subsection {\it 2.2}\/ was expressed in
terms of the variable $u\sim z^2$.  Implementing the change of
variables and remembering the Jacobian factor in \Ftop\ we extract the
prediction
\eqn\nufi{
r_2 = -{g+2\over 6}\ ,
}
which is verified in all of the examples listed below.

\halign{\indent#\qquad\hfil&\hfil#\hfil\cr
&\cr
Example 1:&
$\P4{(1,1,2,2,6)}[12] (2,128)\to \P5{(1,1,1,1,1,3)}[2,6] (1,129)$\cr
(ref.\ \twoparamsI)&$N=2, g=2, r_2=-4/6$\cr
&\cr
Example 2:&
$\P4{(1,1,2,2,2)}[8] (2,86)\to \P5{}[2,6] (1,89)$\cr
(ref.\ \twoparamsI)&$N=2, g=3, r_2=-5/6$\cr
&\cr
Example 3:&
$\P5{(1,1,2,2,2,2)}[4,6] (2,68) \to \P6{}[2,2,3] (1,73)$\cr
(ref.\ \HKTYII)&$N=2, g=4, r_2 =-6/6
$\cr
&\cr
Example 4:&
$\P6{(1,1,2,2,2,2,2)}[4,4,4] (2,58)\to \P7{}[2,2,2,2] (1,65)$\cr
(ref.\ \HKTYII)&$N=2, g=5, r_2 = -7/6$\cr
&\cr
Example 5:&
$\P4{(1,2,2,2,7)}[14] (2,122) \to \P4{(1,1,1,1,4)}[8] (1,149)$\cr
(ref.\ \HKTYII)&$N=2, g=15, r_2=-17/6$\cr
&\cr
Example 6:&
$\P4{(1,2,3,3,3)}[12] (3,69)\to \P5{(1,1,1,1,1,2)}[3,4] (1,79)$\cr
(sec.~4)&$N=3, g=3, r_2=-5/6$\cr
}

\noindent
In the first line of each example, the transition
from Coulomb to Higgs branch is described geometrically.  In the
second line we give the indices $N$ (labeling the gauge group), $g$
(the genus of the singular curve labeling the number of matter
hypermultiplets) and $r_2$, the coefficient the logarithmic term in
$F_1^{\rm top}$.  We also give the relevant reference for the
computation.  The explicit calculations for the first example (which has
a conjectured heterotic dual) and the last example (which has an
$SU(3)$ enhanced symmetry) are in the next section.

\newsec{Two Examples}

In this section we discuss in detail the first and last examples in the list
above.  The first of these was chosen because of the role that it plays in
heterotic/type II duality \refs{\kv,\fhsv}: there is a
conjecture \kv\
that this example is dual to a particular compactification of heterotic string
theory.  Enhanced gauge symmetry in the context of string duality was
discussed in \refs{\kv,\rKKLMV,\paul} for the case in which the point
of enhanced
symmetry lies at weak coupling for the heterotic model.  As we shall
see, this is not the case for the transition we study here.

\subsec{An Example with a Heterotic Dual}

We study type IIA string theory compactified on the Calabi--Yau
hypersurface $M$ in weighted projective space $\P4{(1,1,2,2,6)}$, or
equivalently type IIB string theory on its mirror $W$.  The massless
spectrum of the model contains two massless vector multiplets and 129
massless hypermultiplets, and the moduli space of vacua is
parameterized by the expectation values of the scalars in these
multiplets. By the nonrenormalization theorem of
\refs{\andy,\BBS} the geometry of the scalars in the vector multiplets can be
computed at string tree level. This model has
been extensively studied; in particular, the moduli space of the
scalars in the two vector multiplets is well understood.
Of special importance to us is the singular locus.  This has two
components, one of which was studied in detail (at a particular
limiting point) in \kv .  We will concentrate on the other
component.

This is described as follows.  $M$ is given by the vanishing locus of
a quasihomogeneous polynomial of degree twelve. The mirror manifold $W$
is given at one point in moduli space by a $\IZ_6^2\times\IZ_2$
quotient of $M$.
The most general polynomial of degree twelve invariant under the
quotient is of the form
\eqn\w{
P = x_1^{12}+x_2^{12}+x_3^6+x_4^6+x_5^2 - 12\psi x_1x_2x_3x_4x_5 - 4\phi
x_1^6x_2^6\ .
}
The $\IZ_6^2\times\IZ_2$ quotient thus leads to a two-parameter family
of complex
structures on the mirror $W$, in this example mapping out the entire
space.  The singularity in question arises when $\phi^2=1/4$, for
which the polynomial is singular at the four points
$(1,\omega^j,0,0,0)$ where $\omega^6=1$. These are identified under
the quotient leading to one singular point on the quotient. The
quotient, of course, generates orbifold singularities in the ambient
space, and the smooth mirror model can be obtained by toric blowups of
these. The singularity in question lies at the intersection of several
such orbifold loci, and we need to study the toric resolution.  To
this end we add two new rays to the fan defining the ambient space as
a toric variety.  This means we add two new homogeneous coordinates
and two new $\IC^*$ identifications, in the process introducing two
exceptional divisors to resolve the singularity.

Explicitly, the polytope $\Delta$ may be taken to have vertices
$$(-1,-1,-1,-1),(11,-1,-1,-1),(-1,5,-1,-1),$$
$$(-1,-1,5,-1),(-1,-1,-1,1)$$
and the polar polytope $\D^\circ$ has vertices
$$(-1,-2,-2,-6),(1,0,0,0),(0,1,0,0),$$
$$(0,0,1,0),(0,0,0,1).$$
The face $\Delta_2$ is spanned by the vectors
$(-1,5,-1,-1),(-1,-1,5,-1)$, and $(-1,-1,-1,1)$ while the edge
$\Delta_1^*$ is spanned by $(-1,-2,-2,-6),(1,0,0,0)$.  The face
$\Delta_2$ has the two interior points
$(-1,1,0,0),(-1,0,1,0)$ while $\Delta_2^*$ has the unique
interior point $(0,-1,-1,-3)$.  Thus $g=2$, and $V$ contains an irreducible
exceptional divisor resolving a genus~2 curve of $A_1$ singularities.

To obtain a local description near the singularity we
focus on the four-dimensional cone spanned by the points
$$(-1,-1,-1,-1),(-1,1,0,0),(-1,0,1,0),(-1,-1,-1,1)$$
of $\Delta$, which gives coordinates which we call $(x_1,s,t,x_5)$
($s$ and $t$ correspond to the variables that were called $x_1$ and
$x_2$ in~\Veqn and~\genVeqn.  $x_1$ and $x_5$ correspond to the
similarly named variables of the weighted projective space).
This brings $P$ to the form
\eqn\localform{
P = x_1^{12}-4\phi x_1^6+1 +st(s+t+x_5^2-12\psi x_1x_5)\ .
}
We can now see explicitly the two non-toric divisor classes, in that
the intersection of $W$ with each of the two exceptional divisors has
two components (given by two solutions of the quadratic equation to
which $P$ reduces as discussed above) which coalesce at the
singularity.

We see that at $\phi^2=1/4$, the model is singular along a reducible
curve.  Making the choice $\phi=1/2$, we see that the singular curve
consists of the following rational components, all contained
in the hypersurface $x_1=1$:
\eqn\tet{\eqalign{
s &= t= 0\cr
s &= s+t+x_5^2-12\psi x_5 = 0\cr
t &= s+t+x_5^2-12\psi x_5 = 0\ .\cr
}}
These components intersect at the two points
$$(x_1,s,t,x_5)=(1,0,0,0), (1,0,0,12\psi)\ . $$

The geometry of $M$ is simpler.  The weighted projective space
$\P4{(1,1,2,2,6)}$ is singular along the surface $x_1=x_2=0$.  Blowing it
up yields a toric variety whose exceptional divisor (corresponding to the
edge spanned by $(0,-1,-1,-3)$ above) is a
$\P1{}$ bundle over the weighted projective space
$\P2{(2,2,6)}=\P2{(1,1,3)}$.  After restricting to the exceptional divisor, the
equation for $M$ reduces to an equation in the coordinates of $\P2{(1,1,3)}$
which is quasi-homogeneous of degree~6.  Its zero locus is a
genus $g=2$ curve of $A_1$ singularities when identified with the surface
$x_1=x_2=0$ in $\P4{(1,1,2,2,6)}$, and also gives the equation of a $\P1{}$
bundle over a genus~2 curve in the blowup.

\bigskip\bigskip\noindent{\it Monodromy}\nobreak\bigskip

The prediction about integer monodromy can be verified directly on the
IIB side in this example, and in the related example $\P4{(1,1,2,2,2)}$.
(Each of these
 models has two components to the discriminant divisor; our transition
occurs at the non-principal component.)
In general, such verifications are very difficult because it is hard to
compute integral bases for cohomology.

The $\IZ$-monodromy matrices about the non-principal component for
both $\P4{(1,1,2,2,6)}$ and $\P4{(1,1,2,2,2)}$
are computed in \twoparamsI;
the monodromy matrix for our present example $\P4{(1,1,2,2,6)}$
is in \rKKLMV\ as well
using different bases.
We note that in each case, the monodromy matrix has eigenvalues
$\pm 1$.  The generalized $(+1)$-eigenspace is four-dimensional, and the
generalized $(-1)$-eigenspace is two-dimensional.  We choose a $\IZ$ basis
for each of these generalized eigenspaces, and then take the union of
these two bases to get a basis for the full cohomology.\foot{The resulting
basis is no longer a $\IZ$ basis, but this is not relevant for our
comparison with gauge theory.}\
In this new basis, the monodromy matrix is
\eqn\monomat{\pmatrix{
        1 & 0 & 0 & 0 & 0 & 0 \cr
        0 & 1 & 0 & 0 & 0 & 0 \cr
        0 & 0 & 1 & 0 & 0 & 0 \cr
        0 & 0 & 0 & 1 & 0 & 0 \cr
        0 & 0 & 0 & 0 & -1 & 2(g-1) \cr
        0 & 0 & 0 & 0 & 0 & -1 \cr}}
in both the $\P4{(1,1,2,2,6)}$ and $\P4{(1,1,2,2,2)}$ models, with $g=2$ in the
first and $g=3$ in the second. This is
in complete agreement with our gauge-theoretic prediction~\emon.

\bigskip\bigskip\noindent{\it The Heterotic Dual}\nobreak\bigskip

In \kv\ evidence was given that this model is dual to a heterotic
string theory obtained from a $K3\times T^2$ compactification where
$T^2$ has $\tau=\rho$ by Higgsing all of the nonabelian gauge
symmetries. The two scalars in vector multiplets were identified in
the heterotic theory as the modulus $\tau$ of $T^2$ and the dilaton $S$.
A study of the (other component of the) discriminant locus at a
particular limiting point led to an identification
\eqn\shamit{\eqalign{
j(\tau) &= {(12\psi)^6\over 2\phi}\cr
e^S &= \phi^2\ ,\cr
}}
where $j(\tau)$ is the modular invariant.  These were shown to hold to
first order about the weak-coupling limit $S=-\infty$.  In
\refs{\kv,\pauljan}
it is argued that the exact form of this is given by equating $S,\tau$ with
the {\it flat\/} coordinate approximating the RHS of \shamit\ at
large-$S$.  These identifications locate the locus of enhanced
symmetry $\phi^2=1$ at strong coupling in the heterotic model.

In any event, the enhanced gauge symmetry we observe, as well as the
new branch of hypermultiplet moduli space corresponding to our Higgs
branch have no analog in perturbative heterotic string theory.  This
appears at first sight to contradict the nonrenormalization theorem of
\refs{\andy,\BBS} which in the heterotic string case states that the
hypermultiplet moduli space should be independent of the string
coupling.  This contradiction is resolved if far out along the new
branch we do not find weakly coupled heterotic strings.  We are
thus being given a glimpse of strong-coupling effects in the heterotic
theory. We do not, at this point, have an interpretation of these
effects directly in the heterotic model.  In particular, we do not
know how to represent the light states as heterotic solitons.

\subsec{An Example with $SU(3)$ Gauge Symmetry}

In this subsection, we summarize the calculations for the example of
$\P4{(1,2,3,3,3)}[12]$, which has $SU(3)$ enhanced gauge symmetry.

This weighted projective space has $A_2$ singularities along the
$\P2{}$ defined by $x_1=x_2=0$.  We can blow up this $\P2{}$ torically,
obtaining a toric variety whose fan has edges are spanned by
$$ (-2, -3, -3, -3), (1, 0, 0, 0), (0, 1, 0, 0), (0, 0, 1, 0),$$
$$ (0, 0, 0, 1), (-1, -2, -2, -2), (0, -1, -1, -1).$$
The last two edges correspond to the exceptional divisors of the blowup.

A degree~12 hypersurface intersects our $\P2{}$ in a degree~4 curve
of genus $g=3$.  Correspondingly, the Calabi--Yau hypersurface $M$ contains
two ruled surfaces over a genus~3 curve.
We compute that $h^{1,1}(M)=3$ and $h^{2,1}(M)=69$.  We then compute the
Mori cone of $M$.  In the notation of \HKTYII, it is generated by
\eqn\Morigen{
(-4,-1,0,1,1,1,2,0),(0,1,0,0,0,0,-2,1),(0,0,1,0,0,0,1,-2).
}
The last two generators are the classes of the fibers the two exceptional
divisors.

The program {\sc instanton} computes the Gromov--Witten invariants as

\halign{#\quad\hfil &#\quad\hfil &#\quad\hfil &#\hfil\cr
 $n_{0, 0, 1} = 4$& $n_{0, 0, 2} = 0$& $n_{0, 0, 3} = 0$& $n_{0, 0, 4} = 0$\cr
 $n_{0, 0, 5} =0$& $n_{0, 0, 6} = 0$& $n_{0, 1, 0} = 4$& $n_{0, 1, 1} = 4$\cr
 $n_{0, 2, 0} = 0$& $n_{0, 2, 2} =0$& $n_{0, 3, 0} = 0$& $n_{0, 3, 3} = 0$\cr
 $n_{0, 4, 0} = 0$& $n_{0, 5, 0} = 0$& $n_{0, 6, 0} =0$& $n_{1, 0, 0} = 56$\cr
 $n_{1, 1, 0} = 592$& $n_{1, 1, 1} = 592$& $n_{1, 2, 0} =56$& $n_{1, 2, 1} =
592$\cr
 $n_{1, 2, 2} = 56$& $n_{2, 0, 0} = -272$& $n_{2, 1, 0} =552$& $n_{2, 1, 1} =
552$\cr
 $n_{2, 2, 0} = 9144$& $n_{2, 2, 1} = 64544$& $n_{2, 2, 2} =9144$& $n_{2, 3, 0}
= 552$\cr
 $n_{2, 3, 1} = 64544$& $n_{2, 4, 0} = -272$& $n_{3, 0, 0} =3240$& $n_{3, 1, 0}
= -9088$\cr
 $n_{3, 1, 1} = -9088$& $n_{3, 2, 0} = 15440$& $n_{3, 2, 1} =101664$& $n_{3, 3,
0} = 271824$\cr
 $n_{4, 0, 0} = -58432$& $n_{4, 1, 0} =200268$& $n_{4, 1, 1} = 200268$& $n_{4,
2, 0} = -372416$\cr
 $n_{5, 0, 0} =1303840$& $n_{5, 1, 0} = -5266176$& $n_{6, 0, 0} = -33255216$
&\cr}

\bigskip\noindent
Note that $n_{0,1,0}=n_{0,0,1}=n_{0,1,1}=4=2g-2$, which reflects the
$SU(3)$ gauge symmetry in the way that we discussed earlier.  Similarly,
several of the instanton numbers above are seen to occur in triplets.

We next turn to the calculation of $F_1$.  We do this using the general
strategy described earlier.\foot{We are grateful to A.~Klemm for sharing some
computer code with us which we used to check our calculations.}

We choose coordinates $z_1,z_2,z_3$ near the large complex structure
limit, in the order dictated by our choice of the Mori cone.  The
discriminant has two components, which may be computed by the method of
\refs{\rKap,\MorrisonPlesser}  to be
\eqn\discrims{
\eqalign{
\Delta_1={}&1+64z_1-768z_1z_2-32768z_1^2z_2+196608z_1^2z_2^2
+4194304z_1^3z_2^2\cr&-16777216z_1^3z_2^3
    -294912z_1^2z_2^2z_3-16777216z_1^3z_2^2z_3\cr&+75497472z_1^3z_2^3z_3
    -113246208z_1^3z_2^4z_3^2 \cr
\Delta_2={}&1-4z_2-4z_3+18z_2z_3-27z_2^2z_3^2.\cr}
}
Since $\chi(M)=-132$, we then have by \Ftop
\eqn\Fone{
F_1^{\rm top}=\log({1\over\varpi_0^{17}}\Delta_1^{r_1}\Delta_2^{r_2}
z_1^{s_1}z_2^{s_2}z_3^{s_3}).
}
for some undetermined exponents $r_i,s_j$.
{}Using this expression, the asymptotics which can be calculated
from $c_2(M)$, and the non-existence of certain elliptic curves
we can infer the values
\eqn\exps{
(r_1,r_2)=(-{1\over6},-{5\over6}),\
(s_1,s_2,s_3)=(-5,-{17\over6},-3).
}
The exponent $r_2$ agrees with our gauge-theoretic prediction.

The transition occurs when the exceptional divisors get blown down.  The
monomials which effect the blow down must intersect the last two generators
of \Morigen\ trivially.  These are given by the monomials
\eqn\transmon{
x_1^3x_6^2x_7,x_1x_2x_6x_7,x_3,x_4,x_5,x_2^3x_6x_7^2.
}
In the above, the $x_i$ are the homogeneous coordinates of the toric variety.
Assigning new variables $y_1,\ldots,y_6$ to the respective monomials
in \transmon, we see that
the identity $y_1y_6=y_2^3$ holds.  By examining the weights, we thus see that
we reach a complete intersection $\P5{(1,1,1,1,1,2)}[3,4]$.  Its Hodge
numbers are calculated to be $h^{1,1}=1$ and $h^{2,1}=79$, as anticipated
by \Hodgechange.

\bigbreak\bigskip\bigskip

\noindent
{\bf Note added:} As this work was being readied for publication
we received a preprint \km\ which overlaps this paper to a certain
extent.  It contains two examples not considered here, but which
further confirm our results (the genus is not calculated in \km):

\halign{\indent#\qquad\hfil&\hfil#\hfil\cr
&\cr
Ex.~7 of \km&
$\P4{(1,2,2,3,4)}[12](2,74)\to\P5{(1,1,1,2,2,3)}[6,4](1,79)$\cr
&$N=2,g=4,r_2=-6/6$\cr
&\cr
Ex.~11 of \km&
$\P4{(1,2,3,3,9)}[18](3,99)\to \P5{(1,1,1,1,2,3)}[3,6](1,103)$\cr
&$N=3,g=2,r_2=-4/6$\cr
}

\bigbreak\bigskip\bigskip\centerline{{\bf Acknowledgments}}\nobreak
We thank Brian Greene for collaboration in the early stages of this
work.  We also thank P. Aspinwall, P. Berglund, P. Candelas, A. Font,
M. Gross, A. Klemm, D. Kutasov, A. Schwimmer,
N. Seiberg, C. Vafa, and especially Y. Oz for
discussions, and N. Seiberg for comments on an early version of the
manuscript.
D.R.M. would like to thank the Center for Basic Interactions and the
Einstein Center at the Weizmann
Institute of Science for hospitality during the final stages of
this project.
The work of S.K. was supported in part by NSF grant
DMS-9311386, the
work of D.R.M. was supported in part by NSF grant DMS-9401447,
and the work of M.R.P. was supported by the Center for Basic
Interactions.

\listrefs

\bye